\documentclass[lettersize,journal]{IEEEtran}
\IEEEoverridecommandlockouts
\usepackage{cite}
\usepackage{amsmath,amssymb,amsthm,amsfonts}
\usepackage{algorithm}
\usepackage{algpseudocode}
\usepackage{graphicx}
\usepackage{subfigure}
\usepackage{textcomp}
\usepackage{xcolor}
\usepackage{stfloats}
\usepackage{mathtools}
\usepackage{diagbox}

\def\BibTeX{{\rm B\kern-.05em{\sc i\kern-.025em b}\kern-.08em
    T\kern-.1667em\lower.7ex\hbox{E}\kern-.125emX}}
\graphicspath{{Figures/}{./}}
\usepackage[numbers,sort&compress]{natbib}

\renewcommand{\baselinestretch}{0.9541}

\usepackage{array}
\usepackage{verbatim}
\usepackage{makecell}
\usepackage{multirow}

\usepackage{pifont}
\newcommand{\cmark}{\color{green} \ding{51}}%
\newcommand{\xmark}{\color{red} \ding{55}}%

\begin{document}

\algnewcommand\algorithmicswitch{\textbf{switch}}
\algnewcommand\algorithmiccase{\textbf{case}}
\algnewcommand\algorithmicassert{\texttt{assert}}
\algnewcommand\Assert[1]{\State \algorithmicassert(#1)}%
\algdef{SE}[SWITCH]{Switch}{EndSwitch}[1]{\algorithmicswitch\ #1\ \algorithmicdo}{\algorithmicend\ \algorithmicswitch}%
\algdef{SE}[CASE]{Case}{EndCase}[1]{\algorithmiccase\ #1}{\algorithmicend\ \algorithmiccase}%
\algtext*{EndCase}%

\title{UnifSrv: AP Selection for Achieving \\ Uniformly Good Performance of CF-mMIMO \\ in Realistic Urban Networks


\thanks{The authors are with Chair of Mobile Communications and Computing at RWTH Aachen University, Aachen 52072, Germany (email: \{yunlu.xiao, petrova, simic\}@mcc.rwth-aachen.de).}}

\author{\IEEEauthorblockN{Yunlu~Xiao, Marina Petrova,
and
Ljiljana~Simić}}

\maketitle

\begin{abstract}
Under the ideal assumption of uniform propagation, cell-free massive MIMO (\mbox{CF-mMIMO}) provides uniformly high throughput over the network by effectively surrounding each user with its serving access point (AP) set. However, in realistic \mbox{non-uniform} urban propagation environments, it is difficult to consistently select good limited serving AP sets, resulting in significantly degraded throughput, especially for the \mbox{worst-served} (formerly ``cell-edge'') users. To restore the uniformly good performance of scalable \mbox{CF-mMIMO} in realistic urban networks, we formulate a novel \mbox{multi-objective} optimization problem to jointly achieve high throughput by maximizing the sum data rate, uniform throughput by maximizing Jain's fairness index of the throughput per user, and scalability by minimizing the serving AP set size. We then propose the \textit{UnifSrv} AP selection algorithms to solve this optimization problem, consisting of a deep reinforcement learning (DRL)-based algorithm \mbox{\textit{UnifSrv-DRL}} and a heuristic algorithm \textit{UnifSrv-heu}. We conduct a comprehensive performance evaluation of scalable CF-mMIMO under realistic urban network distributions, propagation, and mobility patterns. Our results show that \textit{UnifSrv} significantly outperforms the prior benchmark AP selection schemes, and for the first time achieves uniformly high throughput of CF-mMIMO under non-uniform urban propagation. Importantly, our heuristic algorithm achieves equivalent throughput to our DRL one, but with orders of magnitude lower complexity. We thus for the first time propose a practical AP selection algorithm that makes CF-mMIMO viable in realistic urban networks. \looseness=-1

\end{abstract}


\section{Introduction}

Cell-free massive multiple-input-multiple-output (\mbox{CF-mMIMO}) has been proposed \cite{CFfirst} as a novel disruptive architecture for 6G mobile networks \cite{CFmag2026}. Different from the traditional cellular network, where a UE is only served by one base station \cite{scboss}, CF-mMIMO deploys dense APs over the network to form a distributed MIMO array to serve all UEs. In this case, all UEs in the network achieve significantly \textit{higher} throughput compared to the cellular network due to the diversity gain of massive MIMO and \textit{more uniform} throughput due to the removal of cell boundaries and the cell-edge effect for the worst-served UEs \cite{cfbook}. CF-mMIMO thus promises to deliver ubiquitous 6G connectivity with ``uniformly great service for everyone'' \cite{CFmag2026, CFfirst}. However, managing the entire network to simultaneously serve a given UE results in impractical signaling overhead and system costs. It is thus vital for CF-mMIMO to reduce the overheads and still preserve the high and uniform throughput by selecting a subset of APs with the best channel conditions to serve a given UE \cite{scalableFirst}.
  
Many algorithms have been proposed \cite{EE, DRLmaxRate, DRLmaxMinRate,  cfsim, GMM, DRLreduceConnect, MAcf, PUCconstDRL} to effectively select this best serving subset for each UE. However, these works make the ideal assumptions of a random uniform distribution of APs and UEs (according to a Poisson point process (PPP)) and a distance-dependent path loss model, whereby the channel conditions over the network are by default assumed to be uniformly distributed. Therefore, when all UEs are served by APs with the best channel conditions to them - corresponding to the UE being effectively surrounded by its AP serving set - their throughput performance will be \textit{uniformly good}, as envisioned in original CF-mMIMO. However, we point out in \cite{mywcnc25} that the layout of realistic network deployments and the associated propagation environments in urban areas are highly non-uniform, leading to a significant degradation in the throughput performance of \mbox{CF-mMIMO}. Under realistic propagation, the channel conditions are generally worse than in an ideal uniform network due to building shadowing. In this case, the existing AP selection algorithms that aim at maximizing the throughput \cite{EE, DRLmaxRate, DRLmaxMinRate} will select a very large number of serving APs to achieve the required serving set channel quality; this results in poor scalability, i.e., a very large serving set size and thus high system cost. On the other hand, the algorithms aiming to form a limited serving AP set that surrounds the UE \cite{cfsim, GMM, DRLreduceConnect} will obtain significantly lower throughput due to bad serving set quality. Finally, the algorithm in \cite{MAcf} that aims at both high throughput and limited serving set size by reducing the \mbox{AP-UE} connections with bad channels will trade off the throughput of the \mbox{worst-served} UEs for a small serving set size. This leads to a non-uniform throughput distribution over the network, in contrast to the design goal of CF-mMIMO.

Therefore, it is vital to reconsider the problem of AP selection for realistic urban environments, as a fundamental prerequisite for making CF-mMIMO work in practice. Importantly, the uniformly high throughput promised by \mbox{CF-mMIMO} over the whole network should not be assumed by default in the highly \mbox{non-uniform} urban propagation environments. Instead, besides the throughput and system cost, the \textit{fairness} in throughput performance among UEs should be explicitly studied and optimized during the AP selection for CF-mMIMO in realistic urban environments. However, to the best of our knowledge, realistic network distributions and propagation are thus far unaddressed in AP selection for \mbox{CF-mMIMO}, putting into question whether \mbox{CF-mMIMO} can in fact achieve both high and uniform throughput in practical mobile networks.

To address this important gap, we present a comprehensive study of the performance of \mbox{CF-mMIMO} under realistic urban propagation environments and show that the existing AP selection methods in \cite{EE, DRLmaxRate, DRLmaxMinRate, MAcf, DRLreduceConnect, cfsim, GMM, DRLmobi} in fact fail to achieve the CF-mMIMO promise of uniformly high throughput. To provide a solution for \mbox{CF-mMIMO} in realistic urban networks, we introduce Jain's fairness index to quantify the uniformity of throughput over the network. We then formulate a novel multi-objective optimization problem to jointly maximize the throughput, maximize the fairness index, and minimize the AP-UE connections. We propose the solution to this optimization problem as \textit{UnifSrv}, two novel serving AP selection algorithms that achieve both high and uniform throughput under highly \mbox{non-uniform} urban propagation: a deep reinforcement learning (DRL)-based algorithm \mbox{\textit{UnifSrv-DRL}} and a low-complexity heuristic algorithm \textit{UnifSrv-heu}. \looseness=-1

Our key contributions are: 

\begin{itemize}

\item We conduct a comprehensive performance analysis of \mbox{CF-mMIMO} under realistic urban network distributions and UE mobility patterns, in realistic propagation environments as given by a site-specific raytracing channel in Seoul and Frankfurt. Our results show that the benchmark AP selection schemes for CF-mMIMO obtain significantly worse throughput in realistic urban environments than in ideal PPP networks, especially for the worst-served UEs, and thus fail to provide uniformly good performance of CF-mMIMO in practice.

\item We formulate and solve a novel multi-objective optimization problem to design \textit{UnifSrv}, AP selection algorithms that achieve \textit{uniformly good} throughput for \mbox{CF-mMIMO} under realistic urban propagation. We maximize both the total throughput and Jain's fairness index of the throughput per UE, and minimize the number of AP-UE connections. We also constrain both the serving AP set size per UE and the number of served UEs per AP to achieve scalability in practical networks.

\item We propose the \textit{UnifSrv-DRL} algorithm by designing a deep Q-learning (DQN) framework to solve the transformed multi-objective optimization problem. We then propose a heuristic algorithm \mbox{\textit{UnifSrv-heu}} by introducing a dynamic fairness-index-based threshold to control the serving AP set expansion. Both of our \textit{UnifSrv} algorithms significantly outperform all considered existing AP selection algorithms and successfully achieve uniformly high throughput for scalable \mbox{CF-mMIMO} under realistic propagation. \looseness=-1

\item Importantly, our heuristic algorithm \textit{UnifSrv-heu} achieves equivalently good throughput performance as \mbox{\textit{UnifSrv-DRL}}, but with orders of magnitude lower algorithm complexity. We thus for the first time propose an AP selection algorithm that achieves both high and uniform throughput of scalable CF-mMIMO in realistic urban environments with low complexity, making CF-mMIMO viable for practical mobile networks. 

\end{itemize}
The rest of this paper is organized as follows: Sec. \ref{related} reviews the related work on AP selection in CF-mMIMO, \mbox{Sec. \ref{model}} details the system model, Sec. \ref{APselect} proposes our \textit{UnifSrv} algorithms, Sec. \ref{results} evaluates the performance of our algorithms and other benchmarks, and Sec. \ref{conclude} concludes the paper.

\section{Related Work}
\label{related}

The performance of CF-mMIMO under realistic urban environments is largely unaddressed in the existing literature, with only a few exceptions \cite{poster, SoftHOcf, Cfmeasure, 2022Evaluation, mywcnc25, drlJP}. The works in \cite{poster, SoftHOcf} take into account the \mbox{site-specific} AP locations in the cities of Seoul and Frankfurt to show the realistic AP location itself already has an impact on the throughput of CF-mMIMO. However, they still model the path loss by a log-distance model and do not consider the impact of realistic propagation. The work in \cite{Cfmeasure} studies a site-specific channel corresponding to \mbox{CF-mMIMO} systems via measurements and \cite{2022Evaluation} studies realistic urban propagation of a mm-wave system via raytracing, showing that the CF-mMIMO architecture indeed achieves channel hardening in realistic multipath propagation. However, neither of them considers any network-wide throughput performance analysis or AP selection. Our work in \cite{mywcnc25} studies the significant throughput performance difference of scalable CF-mMIMO between the \mbox{widely-assumed} log-distance path loss model and the realistic raytracing propagation model, assuming the simple AP selection method from \cite{cfsim}, but without providing solutions to close the performance gap. The work in \cite{drlJP} considers a CF-mMIMO network in an area of Tokyo with a raytracing channel model and mobile UEs and proposes a DRL-based AP selection. However, this algorithm solely aims at reducing the signaling and computational complexity, while the achieved throughput is much lower than the fixed-size AP selection. By contrast, in this paper we for the first time conduct a comprehensive performance analysis of scalable CF-mMIMO using various AP selection schemes in \cite{EE, MAcf, DRLmobi, cfsim} in realistic urban networks, whereby we jointly consider realistic distributions, propagation, and mobility patterns. We then propose two novel AP selection algorithms that achieve uniformly high throughput for scalable CF-mMIMO in realistic urban networks. \looseness=-1

Many works studied serving AP set selection for \mbox{CF-mMIMO} in a static network with a PPP distribution and log-distance path loss model. The works in \cite{EE, DRLmaxRate, DRLmaxMinRate} propose \textit{pure UE-centric (PUC)} algorithms that focus on maximizing the throughput. In \cite{EE} a \mbox{large-scale-fading-based} algorithm is proposed to maximize the throughput per UE by selecting the APs with the highest received signal strength to serve a given UE until the total channel gain of the serving set achieves 95\% of the channel gain where all APs serve the UE. A deep deterministic policy gradient algorithm is proposed in \cite{DRLmaxRate} to both maximize the total throughput over the network and satisfy the minimum required throughput, and therefore especially improve the throughput of the \mbox{poorly-performing} UEs compared to the fixed-size AP selection. Similarly, AP selection based on federated multi-agent reinforcement learning is proposed in \cite{DRLmaxMinRate} to maximize both the sum throughput and the throughput of \mbox{worst-served} UEs. However, these algorithms do not consider the key practical serving AP set size constraint of scalable CF-mMIMO. We show in Sec. \ref{benchmark} that the \textit{PUC} benchmark thus obtains very large serving sets under the realistic urban environment, in order to achieve a high serving set quality with generally poor channels.

\begin{table*}[t]
\caption{Related Work on AP Selection for CF-mMIMO.}
\centering
\footnotesize
\setlength{\tabcolsep}{3.2pt}
\renewcommand{\arraystretch}{1}
\begin{tabular}{l c c c c c} 
{\textbf{Reference}} 
& {\textbf{Maximize throughput}} 
& {\textbf{Constrain AP serving set size}} 
& {\textbf{Model fairness}} 
& {\textbf{Reduce pilot contamination}} 
& {\textbf{Realistic propagation}} \\ 
\hline
\hline

\cite{EE, DRLmaxRate, DRLmaxMinRate}, \textit{PUC} 
& \cmark  
& \xmark 
& \xmark  
& \xmark
& \xmark \\

\cite{cfsim, GMM, DRLreduceConnect}, \textit{CUC} 
& \xmark  
& \cmark 
& \xmark 
& \xmark
& \xmark \\

\cite{MAcf, PUCconstDRL}, \textit{PUC-const}
& \cmark 
& \xmark 
& \xmark 
& \cmark
& \xmark \\

\cite{DRLmobi} (cellular), \textit{PF-DRL}  
& \xmark 
& \cmark  
& \cmark  
& \cmark  
& \xmark  \\


\textbf{This paper, \textit{UnifSrv}}       
& \cmark  
& \cmark  
& \cmark  
& \cmark
& \cmark \\

\end{tabular}
\label{related_work}
\end{table*}

The works in \cite{cfsim, GMM, DRLreduceConnect} consider limiting the serving AP set size. The \textit{clustered \mbox{UE-centric} (CUC)} algorithms are proposed in \cite{cfsim, GMM} to constrain the serving AP set size per UE by assigning the APs to multiple disjoint clusters and then serving the UEs with a limited number of clusters. In particular, a heuristic algorithm is proposed in \cite{cfsim}, where the UEs are served by the few clusters to which the APs with the best channels belong. A Gaussian Mixture Model algorithm is proposed in \cite{GMM} to optimize the serving cluster selection to achieve both high average and 90\%-likely throughput. A DRL-based algorithm is proposed in \cite{DRLreduceConnect} to minimize the \mbox{AP-UE} connections and achieve high throughput with reduced fronthaul requirements. However, these designs assume the channel conditions of all APs in a similar location, e.g., in the same cluster, to be similar; this does not hold under realistic propagation \cite{mywcnc25}. We show in Sec. \ref{benchmark} that these algorithms result in a low throughput under realistic urban propagation.
 
A \textit{constrained pure UE-centric (PUC-const)} algorithm is proposed in \cite{MAcf, PUCconstDRL} to constrain the number of served UEs to the available number of pilots per AP, and therefore both achieve a serving set with good channel conditions for UEs and reduce pilot contamination for APs. A DRL-based algorithm \mbox{\textit{PF-DRL}} is proposed in \cite{DRLmobi} to optimize the proportional fairness of throughput per UE and achieve uniform throughput over a cellular network. However, we show in Sec. \ref{benchmark} that \mbox{\textit{PUC-const}} trades off the throughput of the worst-served UEs to satisfy the constraint, leading to a non-uniform throughput performance. Conversely, \textit{PF-DRL} applied to CF-mMIMO trades off the throughput of the well-served UEs to achieve proportional fairness, resulting in low sum throughput over the network.

Overall, these prior algorithms \cite{EE, DRLmaxRate, DRLmaxMinRate, MAcf, DRLreduceConnect, cfsim, GMM, DRLmobi} consider the objective of maximizing throughput, limiting serving set size, reducing pilot contamination, or achieving uniform throughput \textit{separately}. By contrast, our work \textit{jointly} considers all these objectives and constraints, optimizing the throughput of the \mbox{poorly-performing} UEs by optimizing the throughput performance fairness over the whole network, thus achieving high \textit{and} uniform throughput across the network. Finally we note that the prior algorithms in \cite{EE, DRLmaxRate, DRLmaxMinRate, MAcf, DRLreduceConnect} are based on the impractical \mbox{UE-centric} \mbox{CF-mMIMO} architecture with only one central processing unit (CPU) managing the entire network, while the works in \cite{cfsim, GMM} assume multiple CPUs in the network but only form the serving set with complete CPU clusters. By contrast, we consider the general practical open radio access network (O-RAN) architecture proposed in \cite{ORAN2025} with multiple CPUs in the network, and decouple the serving AP set and the CPU cluster by allowing the UE to be served by a subset of APs in a CPU cluster as in \cite{ORAN2025}. We summarize the related work on AP selection for \mbox{CF-mMIMO} in Table \ref{related_work}. \looseness=-1

\section{System Model}
\label{model}

\subsection{Network Topology \& Mobility Model}
\label{topo}

\begin{figure}[!tb]
   \vspace{0.05in} 
	\centering
	\subfigure[Seoul, 50 tracks]{
		\label{topo.seoul}
		\includegraphics[width=0.31\linewidth]{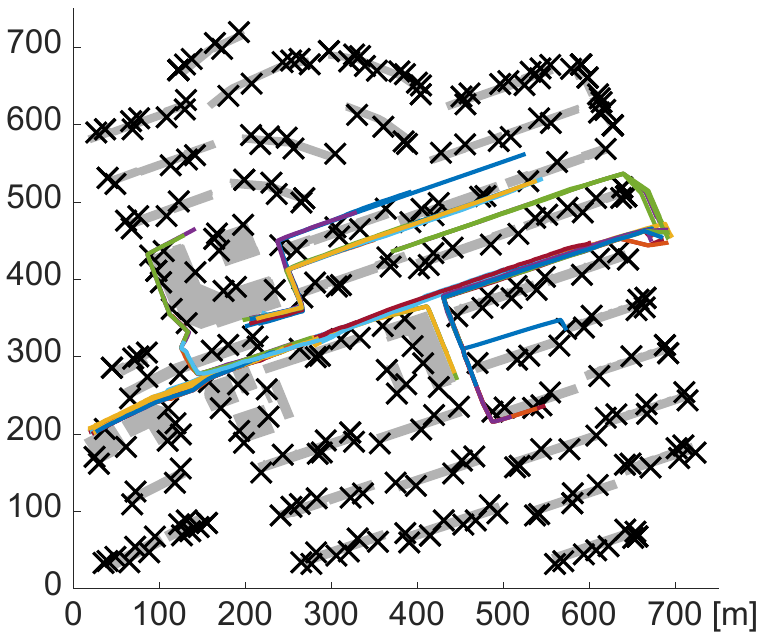}}
	\subfigure[Frankfurt, 50 tracks]{
		\label{topo.frankfurt}
		\includegraphics[width=0.31\linewidth]{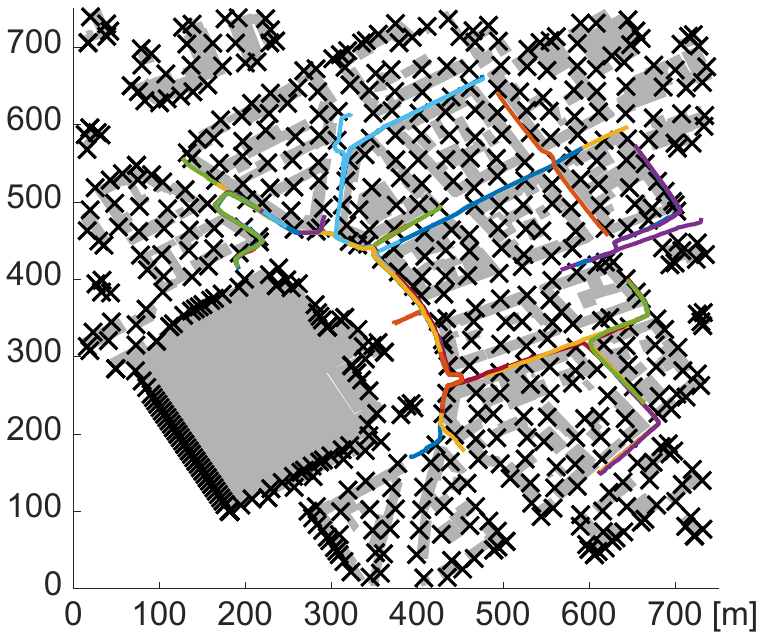}}
	\subfigure[PPP, a RWP track]{
		\label{topo.ppp}
		\includegraphics[width=0.31\linewidth]{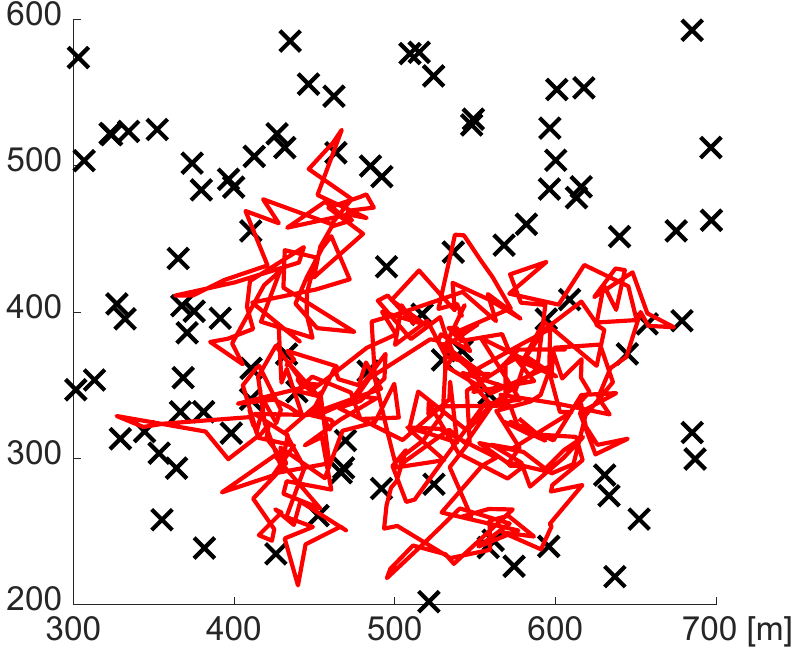}}
	\caption{Network topologies and UE mobility patterns, with buildings marked with gray polygons, AP locations marked with black crosses, and UE tracks marked with colored lines (not all tracks individually visible due to overlap). \looseness=-1}
	\label{topofig}
	\vspace{-0.1cm}
\end{figure}
We consider a network with $M$ static APs (AP set denoted as $\mathcal{M}$) and $K$ mobile UEs ($\mathcal{K}$) distributed in an area \mbox{$S=750 \text{ m} \times 750 \text{ m}$}. We consider two real city areas in Seoul and Frankfurt with qualitatively dissimilar layouts, representing different urban build-up densities, as our study areas. The building layouts are obtained via a public open data base as shapefiles \cite{shp}. We consider $M=324$ APs ($\lambda = 576 \text{ AP/km}^2$) corresponding to a medium-density urban network in Seoul and $M=625$ APs ($\lambda = 1111 \text{ AP/km}^2$) corresponding to a high-density urban network in Frankfurt \cite{poster}. We generate the AP locations via the AP placement algorithm in \cite{mywcnc25}, whereby the APs are distributed as uniformly as possible to best achieve the ideal ``cell-free'' deployment \cite{cfvs}, but we ensure the AP locations to be practical by only placing the APs at edges or corners of the building. We assume $K=50$ ($\lambda_U = 89 \text{ UE/km}^2$) mobile UEs\footnote{We note that the $K \approx 50$ UEs in the network considered in our and prior \mbox{CF-mMIMO} works \cite{scalableFirst, cfbook, EE, DRLmaxRate, DRLmaxMinRate,  cfsim, GMM, DRLreduceConnect, MAcf} in practice corresponds to the active users simultaneously scheduled over a common time-frequency resource.} in the network, leading to a AP/UE ratio of around 6:1 and 12:1 in the Seoul and Frankfurt areas, respectively, in line with prior works \cite{EE, DRLmaxRate, DRLmaxMinRate,  cfsim, GMM, DRLreduceConnect, MAcf} and satisfying the massive MIMO configuration \cite{MIMObook} for sufficient spatial diversity and channel hardening. \looseness=-1

In addition to realistic AP distributions, we also consider practical UE mobility patterns to obtain realistic propagation for these UE locations and also consider channel aging to quantify the throughput under mobility\footnote{We note that mobility management for CF-mMIMO subsequently applies a handover algorithm to decide whether the serving set will be changed to the candidate set given by the AP selection algorithm \cite{DRLmaxMinRate, SoftHOcf, beerten2025}. Handover algorithm design can thus be considered independent of AP selection and is out of the scope of this paper.}.  We assume a pedestrian UE speed of $0.8$ m/s for Frankfurt and a cycling speed of $3.6$ m/s for Seoul to represent typical urban UE mobility. We generate these mobility patterns for each UE via VisWalk \cite{viswalk} for the two city areas based on their building layouts, where individual UE tracks are often very similar. We show in Fig. \ref{topo.seoul} the network topology and mobility patterns of all $K=50$ UEs in Seoul and those in Frankfurt in Fig. \ref{topo.frankfurt}. As a reference, we also consider the scenario with a PPP distribution of the $M$ APs and the random way point (RWP) mobility model \cite{rwp} for the $K$ UEs in the network. Namely, the UEs move with a constant speed same as the corresponding urban networks during each mobility period, with a random direction and Rayleigh distributed transition lengths. We show in Fig. \ref{topo.ppp} an example RWP UE track in a subarea of $S$.

\subsection{CF-mMIMO Network Architecture}
\label{Arch}

\begin{figure}[!tb] 
   \vspace{0.05in} 
	\centering
	\subfigure[]{
		\label{arch}
		\includegraphics[width=0.65\linewidth]{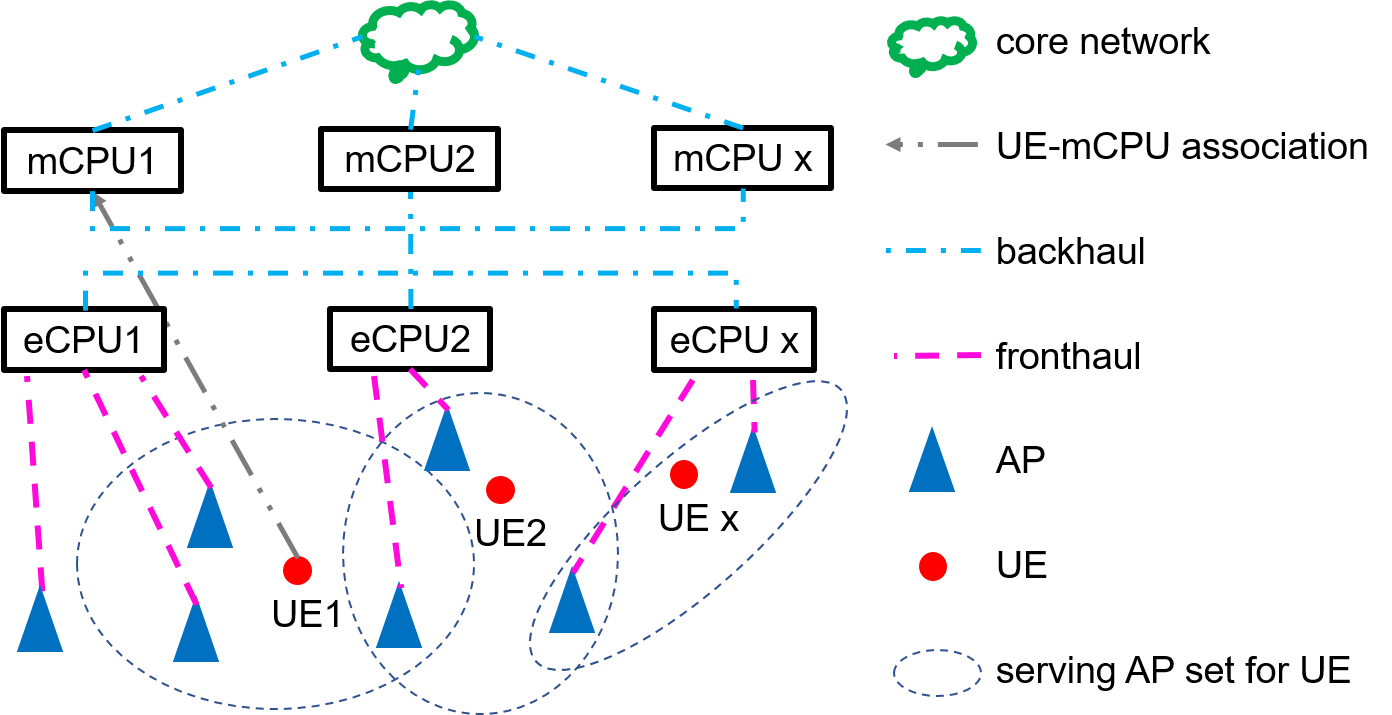}}
	\subfigure[]{
		\label{block}
		\includegraphics[width=0.31\linewidth]{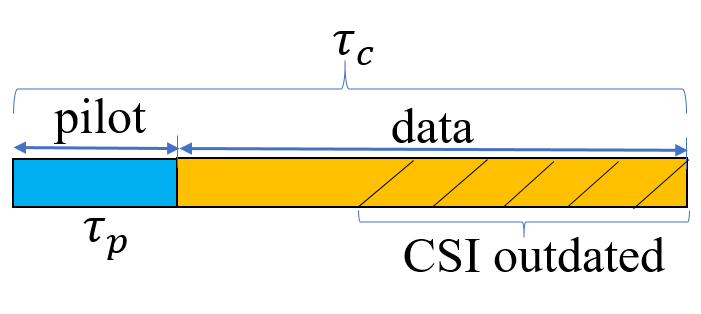}}
	\caption{Illustrations of (a) scalable CF-mMIMO architecture, (b) downlink communication block.}
	\label{archblock}
\end{figure}

We consider the downlink of a scalable CF-mMIMO architecture based on O-RAN as in \cite{ORAN2025}, which allows assigning each UE an arbitrary limited serving AP set. \mbox{Fig. \ref{archblock}} illustrates the network architecture and downlink communication block. The APs are managed in clusters and connected to the edge CPU (eCPU) of the cluster via fronthaul. The eCPUs connect to each other and to the manager CPUs (mCPUs) and eventually to the core network via backhaul. To avoid unlimited signaling and precoding complexity, each UE is served by a subset of APs, which can be managed by multiple eCPUs. However, a UE is only associated with one mCPU for signal processing \cite{ORAN2025}. We adopt the dynamic cooperation matrix in \cite{scalableCF} to describe the connections between all APs and UEs over the network. We define the $M \times K$ cooperation matrix as $D$. If AP $m$ serves UE $k$, $D_{mk} = 1$, otherwise $D_{mk} = 0$. We denote the serving set size per UE $k$ as \mbox{$G_k=\sum^M_{m=1} D_{mk}$} and the number of served UEs per AP $m$ as \mbox{$W_m=\sum^K_{k=1} D_{mk}$}. \looseness=-1

We assume $D$ is known to all CPUs so that each mCPU and eCPU is aware of its associated UEs and APs via $D$. For downlink transmission, the mCPU distributes the UE data to the associated eCPUs of this UE. The eCPUs then preform multi-UE precoding for the served UEs. For example, in \mbox{Fig. \ref{arch}}, UE 1 is served by the APs in eCPU clusters 1 \& 2 and is managed by mCPU 1. The mCPU 1 then distributes the downlink data for UE 1 to eCPUs 1 \& 2, allowing them to calculate precoding vectors and distribute the data to the APs they manage. The APs do not perform any processing, but only act as remote radio heads that wirelessly send the signal from the eCPU to the UE. \looseness=-1

\subsection{Channel Model}
\label{channel}
\begin{table}[!tb]
\caption{Raytracing Simulation Parameters}
\centering
\begin{tabular}{ll}
\hline
\textit{Parameter}              & \textit{Value}                \\ \hline
transmit power         & 20 dBm              \\
carrier frequency      & 2 GHz               \\
bandwith, $B$          & 20 MHz               \\
noise figure           & 9 dB                 \\

antenna                & Half-wave dipole     \\
building material      & ITU Concrete 2.4 GHz \\
propagation model      & X3D raytracing       \\
number of reflections/diffractions  & 6/1                    \\
outage threshold   & -250 dBm            \\ \hline
\end{tabular}
\label{insitepara}
\end{table}
We consider a downlink communication block as in \mbox{Fig. \ref{block}} with a total length of $\tau_c=200$ and assume the channel measurement, AP selection, and signal processing only happen every $\tau_c$ slots. Since the channels are only measured at the beginning of the communication block, channel information is outdated within the block and the channel realizations at different time instants in a block are different due to UE mobility. To model this channel aging effect, we assume the channel realizations at different time instants in a block are correlated as modelled in \cite{aging}. We further assume the correlation coefficient between the channel realizations at time 0 and time instant $t$ is modelled as the zeroth-order Bessel function of the first kind \cite{aging}
\begin{equation}
\rho[t]=J_0\left[2 \pi \frac{v f_c}{c} T_{s} \left(t-\tau_p-1\right)\right],
\label{rho}
\end{equation}
\noindent where $v$ is the speed of the UE, $f_c$ is the carrier frequency, $c$ is the speed of light, $T_{s}$ is the sampling time (i.e., the duration of one slot), and $\tau_p$ is the length of the pilot sequence. The channel between AP $m$ and UE $k$ at time $t$ is thus \cite{aging} \looseness=-1
\begin{equation}
h_{m k}[t]=\rho[t] h_{m k}[0]+\sqrt{1-\rho^2[t]} g_{m k}[t],
\end{equation}
\noindent where $h_{m k}[0]$ is the channel at time 0, i.e., the initial state, and $g_{mk}[t]$ is the component that is independent of channel aging. Assuming Rayleigh fading for the small-scale fading model, we have $h_{mk}[0] \sim N_{C}(0,R_{mk})$ and $g_{mk}[t] \sim N_{C}(0,R_{mk})$, where \mbox{$R_{mk} = 1/L_{mk}$} represents the large-scale fading of the link between AP $m$ and UE $k$ \cite{scalableCF}, where $L_{mk}$ is the path loss.\looseness=-1

We obtain the average path loss $L_{mk}$ in two ways. First, for our realistic urban areas of Seoul and Frankfurt, we obtain the large-scale path loss via raytracing as a deterministic channel model realistically representing the site-specific wireless propagation environment. We obtain the channel data using the Wireless Insite software from RemCom \cite{insite}, based on the city layouts (\textit{cf.} Fig. \ref{topofig}) and the simulation settings in Table \ref{insitepara}. Second, as a reference, we consider the widely adopted Hata-COST231 \mbox{three-slope} \mbox{log-distance} propagation model \cite{cfvs} for the PPP networks as a more simplistic statistical channel model, as typically assumed in the prior CF-mMIMO literature. This model gives the path loss with shadowing between AP $m$ and UE $k$ as $L_{m k} = \overline{\mathrm{L}}_{m k} 10^{\sigma z_{mk} /10}$, with $\sigma$ as the shadowing standard deviation, $z_{mk} \sim N(0,1)$, and the average path loss in dB given by \looseness=-1
\begin{equation}
\overline{\mathrm{L}}_{m k}\!\!=\!\!\left\{\begin{array}{l}
\!\!\!\!L_0\!+\!35 \log _{10}\left(d_{m k}\right)\!, d_{m k}\!>\!d_c \\
\!\!\!\!L_0\!+\!15 \log _{10}\left(d_c\right)\!+\!20 \log _{10}\left(d_0\right)\!, d_{m k} \!\leq\! d_0 , \\
\!\!\!\!L_0\!+\!15 \log _{10}\left(d_c\right)\!+\!20 \log _{10}\left(d_{m k}\right)\!, \text{otherwise} \\
\end{array}\right.
\label{log}
\end{equation}
\noindent where $d_{m k}$ is the distance between AP $m$ and UE $k$, and
\begin{equation}
\begin{aligned}
L_0 \!= & 46.3\!+\!33.9 \log _{10}(f_c)\!-\!13.82 \log _{10}\left(a_{\mathrm{AP}}\right) \\
& \!-\!\left(1.1 \log _{10}(f_c)\!-\!0.7\right) a_{\mathrm{UE}}\!+\!\left(1.56 \log _{10}(f_c)\!-\!0.8\right),
\end{aligned}
\end{equation}
\noindent where $a_{AP}$ is the AP antenna height, $a_{UE}$ is the UE antenna height, $d_c=50$ m, and \mbox{$d_0=10$ m}. \looseness=-1

\subsection{Signal Processing \& Throughput Model}
\label{signal}
For signal processing, we assume a pilot sequence occupying $\tau_p=10$ slots in each communication block, minimum mean-square error (MMSE) channel estimation, equal power allocation, and partial MMSE precoding \cite{scalableCF}. We adopt random pilot assignment, where a shared pool of $\tau_p$ pilots is available network-wide and each AP randomly assigns these pilots to its served UEs. We thus assume no inter-AP pilot contamination as in \cite{cfsim, GMM, DRLreduceConnect}, since random pilot assignment is known to result in low mutual interference of co-pilot UEs \cite{Rpilot}. The received signal at UE $k$ for the downlink at time $t$ is modelled as \looseness=-1
\begin{equation}
x_k [t]\!=\!\!\sum_{m=1}^M \! h_{m k}[t] \! \left(\!\sum_{i=1}^K \! \sqrt{p_{m i}} D_{mi} w_{m i} q_{m i} \!\! \right)\!\! +\! n_0,
\label{signalY}
\end{equation}
\noindent where $p_{mi}$ is the downlink transmit power, \mbox{$q_{mi} \sim N_C(0,1)$} is the downlink transmit signal, and $w_{mi}$ is the precoding vector of AP $m$ for UE $i$ given by (18) in \cite{aging}.  

With the MMSE channel estimation, the channel estimate is given by $\hat{h}_{mk}[t] \sim N_C(0, Z_{mk}[t])$ \cite{aging}, where
\begin{equation}
Z_{m k}[t]=\frac{\rho^2\left[\tau_p+1-t\right] \beta_{m k}^2 n_0}{p_{m k} \sum_{k \in \mathcal{P}_k} \beta_{m k} n_0+p_{m k}},
\label{Qmk}
\end{equation}
\noindent with $\mathcal{P}_k$ as the UE set that uses the same pilot as UE $k$; (\ref{Qmk}) thus models the channel aging effect in channel estimation and consequently the throughput performance as follows.

With centralized partial MMSE precoding, the eCPUs (\textit{cf.} Sec. \ref{Arch}) calculate the precoding vector $w_{mk}$ using the channel estimates $\hat{h}_{mk}[t]$ given by (\ref{Qmk}). The serving eCPUs then precode the signal with $w_{mk}$ to obtain $x_k[t]$. Given the received signal in (\ref{signalY}) and this precoding process, the \mbox{signal-to-interference-and-noise} ratio (SINR) for UE $k$ is given by \cite{aging} \looseness=-1
\begin{equation}
\gamma_k=\frac{\zeta }{I+n_0},
\label{sinr}
\end{equation}
\noindent where
\begin{equation}
\begin{aligned}
& \zeta\!=\!\rho^2[t\!-\!\tau_p\!-\!1] p_{m k}\!\left| \mathbb{E}\left\{ \sum_{m=1}^M h^H_{m k}[t]  D_{m k} w_{m k}\right\}\right|^2 \\
& I=\sum_{i=1}^K p_{m i} \mathbb{E}\left\{ \left| \sum_{m=1}^M h^H_{m k}[t] D_{m i} w_{m i}\right|^2\right\}-S^d
\end{aligned}.
\label{eqsi}
\end{equation}
\indent The throughput of UE $k$ is then given by \cite{ORAN2025}
\begin{equation}
R_k=B \frac{\tau_c-\tau_p}{\tau_c} \log (\operatorname{\gamma}_k+1),
\label{rate}
\end{equation}
\noindent where $B$ is the system bandwidth.

\section{AP Selection for CF-mMIMO Under Realistic Urban Propagation}
\label{APselect}

\subsection{Problem Formulation}
\label{problem}
In the ideal PPP networks with uniform channel conditions, the existing AP selection algorithms in \cite{EE, DRLmaxRate, DRLmaxMinRate, MAcf, DRLreduceConnect, cfsim, GMM, DRLmobi} inherently achieve uniformly good throughput with small serving AP set sizes, even though they consider the objectives of high throughput, uniform performance, or limited serving set size separately. However, under realistic urban propagation, the channel conditions are highly non-uniform and generally worse than in an ideal PPP network \cite{mywcnc25}. We show in \mbox{Sec. \ref{benchmark}} that these prior benchmarks consequently perform poorly in realistic urban networks due to their incomplete objective consideration. Specifically, the \textit{PUC} algorithms \cite{EE, DRLmaxRate, DRLmaxMinRate} maximize only throughput, resulting in very large serving AP sets to improve the serving set quality via AP quantity. The \textit{CUC} algorithms \cite{cfsim, GMM, DRLreduceConnect} only limit the serving set size, leading to low throughput due to poor serving set quality. The \textit{PUC-const} algorithm \cite{MAcf} pursues both high throughput and limited serving set size, but trades off the throughput of the \mbox{worst-served} UEs for a small serving set size, causing \mbox{non-uniform} throughput. The \textit{PF-DRL} algorithm \cite{DRLmobi} maximizes solely the throughput fairness, obtaining uniform but low throughput by trading off the throughput of the \mbox{well-served} UEs to achieve proportional fairness. Therefore, it is necessary to instead \textit{jointly} optimize sum throughput, per UE throughput fairness, and limiting the number of AP - UE connections for AP selection under realistic propagation to achieve uniformly good performance of scalable CF-mMIMO.

To address this gap, we formulate a novel multi-objective optimization problem for AP selection in realistic urban networks as follows. We consider the throughput distribution over all UEs as the primary quality-of-service metric to fulfill the promise of \textit{high and uniform} throughput in CF-mMIMO. Consequently, we obtain high throughput via maximizing the total data rate and uniform throughput via maximizing Jain's fairness index $\Phi'$ of the data rate per UE. We note that Jain's fairness index is a well-established metric \cite{shortFI, optimalFI} for quantifying uniformity of resource allocation. The values of $\Phi'$ lie in $[1/K, 1]$; when $\Phi' \to 1$, the UEs achieve similar throughput over the network and the system is considered to be fair, aligning with the promise of uniform throughput in CF-mMIMO. We also minimize the number of AP - UE connections to reduce the signaling and processing cost \cite{scalableCF}. We constrain the number of UEs served by each AP $W_m$ to $W_{max}=\tau_p$, i.e., no more than the pilot length, to prevent intra-AP pilot contamination \cite{MAcf}. We also constrain the number of APs serving a UE $G_k$ to $G_{max} < M$ to achieve scalability and limited system cost \cite{scalableCF}. The decision variable $D_{mk}$ is a binary variable that states whether AP $m$ is connected to UE $k$ or not. We formulate the optimization problem as: \looseness=-1
\begin{subequations}
\label{eq:optimization}
\begin{align}
\max_{D_{mk}} \;\; & \sum\nolimits_{k=1}^K R_k, 
\label{objSumRate} \\
\max_{D_{mk}} \;\; & \Phi' = \frac{\left( \sum_{k=1}^K R_k \right)^2}{K \sum_{k=1}^K R_k^2},
\label{objFair} \\
\min_{D_{mk}} \;\; & \sum\nolimits_{k=1}^K \sum\nolimits_{m=1}^M D_{mk},
\label{objG} \\
\text{s.t.} \quad 
& W_m=\sum\nolimits_{k=1}^K D_{mk} \leq W_{max}, \quad \forall m \in \mathcal{M},
\label{cstAP} \\
& G_k=\sum\nolimits_{m=1}^M D_{mk} \leq G_{max}, \quad \forall k \in \mathcal{K}, 
\label{cstUE}\\
& D_{mk} \in \{0,1\}, \quad \forall m \in \mathcal{M}, \; \forall k \in \mathcal{K}.
\label{constrain}
\end{align}
\end{subequations}

\subsection{Novel UnifSrv AP Selection Algorithms}
\label{algo}
The optimization problem in (\ref{eq:optimization}) is a non-linear integer programming problem, which is known to be NP-hard. To solve the problem, we first transform (\ref{eq:optimization}) by simplifying the data rate $R_k$. According to (\ref{rate}), the data rate monotonically increases with the SINR $\gamma_k$, and therefore the optimization objectives (\ref{objSumRate}) and (\ref{objFair}) can be transformed to optimize the sum and fairness index of SINR per UE given in (\ref{sinr}), i.e., replace $R_k$ with $\gamma_k$ in (\ref{eq:optimization}). Furthermore, since $\gamma_k$ given by (\ref{eqsi}) depends on the small-scale fading coefficients and the corresponding precoding vectors, its exact evaluation requires full channel state information \textit{after} AP selection. Therefore, $\gamma_k$ is practically infeasible\footnote{Although one could in principle alternatively evaluate the exact SINR over a candidate serving AP set, such an approach requires repeated CSI estimation and precoder computation for all candidate sets, in the order of $O(M^{G_{max}})$, incurring prohibitive complexity in a practically large network.} to be used in AP selection and requires transformation. Under the assumed \mbox{CF-mMIMO} network architecture (\textit{cf. }Sec. \ref{topo}), the number of APs is sufficiently large and the channels across all \mbox{AP - UE} links are assumed to be independent. Due to channel hardening, the aggregated effective channel gain is asymptotically deterministic and the resulting SINR can be accurately approximated as a function depending only on the large-scale fading coefficients \cite{MIMObook}. In this case, the desired signal power is proportional to the aggregate large-scale fading from the serving AP set, i.e., $\sum^M_{m=1} D_{mk} \beta_{mk}$, while the interference is approximated by the remaining large-scale fading contributions from \mbox{non-serving} APs. Consequently, consistent with prior works \cite{scalableCF, aging, EE, DRLmaxRate, DRLmaxMinRate, cfsim, GMM, DRLreduceConnect, MAcf}, we transform the exact SINR in \eqref{sinr} to
\begin{equation}
\gamma_k \approx S_k=\frac{\sum^M_{m=1} D_{mk} \beta_{mk}}{\sum^M_{m=1} \beta_{mk}-\sum^M_{m=1} D_{mk} \beta_{mk}+1},
\label{simpleSINR}
\end{equation}
where $\beta_{mk} = p_{mk} / (L_{mk} n_0)$ is the average SNR of the channel between AP $m$ and UE $k$ over a communication block.

The objective (\ref{objSumRate}) and (\ref{objFair}) can finally be transformed by replacing $R_k$ with $S_k$. We emphasize that (\ref{simpleSINR}) is only used to relax and solve the optimization problem (\ref{eq:optimization}). We still use the full instantaneous SINR in (\ref{sinr}) in our performance evaluation in \mbox{Sec. \ref{results}}. In the following, we propose a DRL approach \mbox{\textit{UnifSrv-DRL}} and a \mbox{threshold-based }heuristic algorithm \textit{UnifSrv-heu} to solve the transformed optimization problem.

\subsubsection{UnifSrv-DRL}
\label{myDRL}

We propose a DRL algorithm to solve (\ref{eq:optimization}) as follows. We model the AP selection of \mbox{CF-mMIMO} as a Markov decision process (MDP), with the state $s$ as the channel condition of all possible AP-UE connections and the action $a$ as all the possible values of the $M \times K$ connection matrix $D$. We deploy one agent at the core network to learn the AP selection policy by interacting with the environment by trial and error and receiving a reward based on its actions via the optimal \mbox{action-value} function $Q(s, a)$. The dimension of both the state and action space grows to $2^{MK}$. To avoid visiting every possible state in such a large state space, we let the agent make the connection decision of each UE one by one, so that the agent needs only to observe the channel state and serving set of a given UE $k$. We then use a neural network to approximate the action-value function $Q_\theta(s, a)$, i.e., we apply the DQN algorithm to perform the DRL. \looseness=-1

\textbf{State space}: The state of UE $k$ is given by \mbox{$s_k=\left\{D_{1 k} \beta_{1 k}, \ldots, D_{M k} \beta_{M k},W_1, \ldots, W_{G_{max}} \right\}$}, which observes the channel quality of the current serving AP set in terms of $\beta_{mk}$ and the serving capacity of these APs in terms of $W_k$. \looseness=-1

\textbf{Action space}: The action space is the set of all APs that provide a sufficient SNR to UE $k$, given by \mbox{$a_k = \{ m \in {1,...,M} | \beta_{mk} \geq \beta_0 \}$}, where $\beta_0$ is the SNR threshold for valid connections. The APs with an SNR below a typical decoding threshold should not be considered as a serving AP to this UE at all. We thus remove these APs from the action space to speed up the training.

\begin{figure}[!tb] 
	\centering
		\includegraphics[width=1\linewidth]{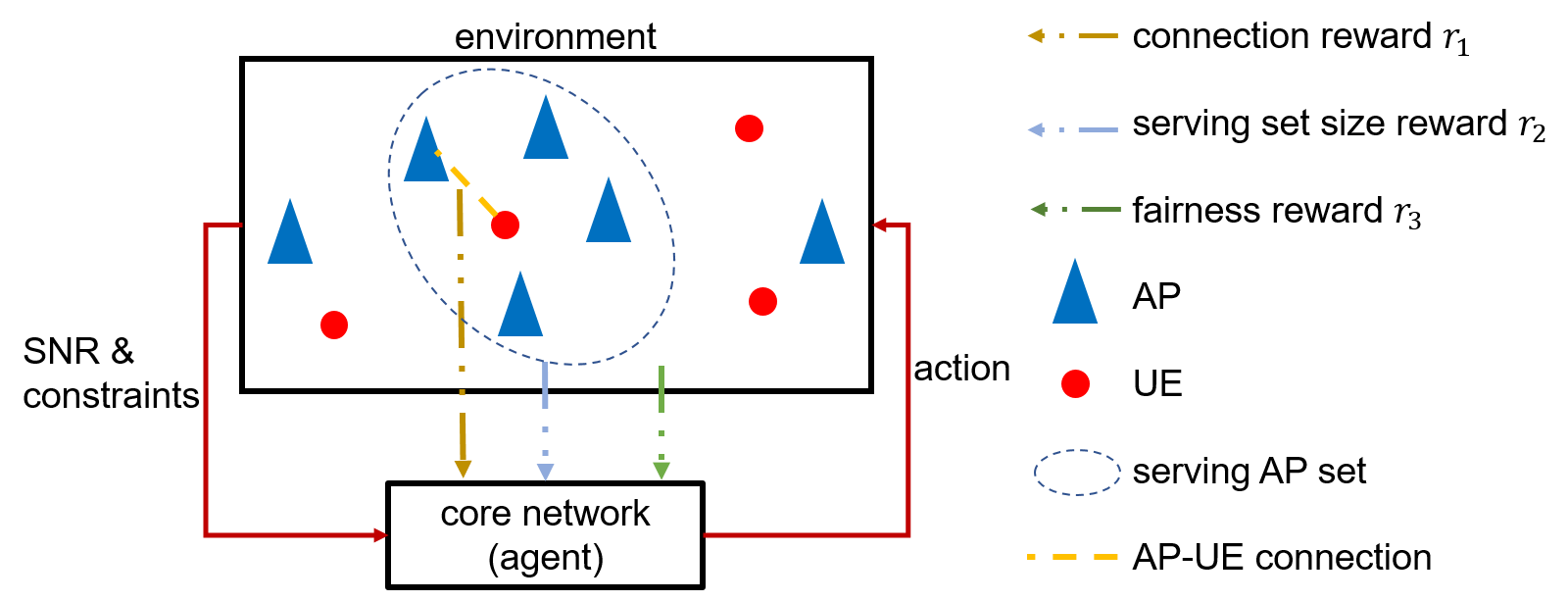}
	\caption{The DRL-based AP selection algorithm framework of \textit{UnifSrv-DRL}. \looseness=-1}
\label{DQN}
\end{figure}

\textbf{DQN algorithm \& reward function}: Fig. \ref{DQN} illustrates the DRL-based AP selection algorithm framework. At each step within an episode, the agent performs an action to connect a UE $k$ to a random AP within the action space. For each step, we offer an immediate reward $r_1 = \omega_1 \beta_{m^\star k} / \beta_{max}$, where $\omega_1$ is a constant representing the weight, $m^\star$ is the index of the AP chosen in this action, and $\beta_{max}$ is the highest SNR this UE can obtain from any AP. This reward encourages the agent to connect UE $k$ with the APs with as good channel quality as possible. We also apply a penalty if the AP chosen cannot serve more UEs, i.e., $r_1 = -1$ if $W_{m^\star} \geq W_{max}$. This teaches the agent to satisfy the constraint in (\ref{cstAP}). After $U_m$ steps or when the number of candidate serving APs for this UE reaches $G_{max}$, which we consider as one round, the agent moves on to the next UE. In this case, each UE is served by at most $G_{max}$ APs, which satisfies the constraint in (\ref{cstUE}). For each round, we design an intermediate reward function to evaluate the serving set size, given by
\begin{equation}
r_2=\omega_2 (1-\sum_{m=1}^M D_{m k}/M),
\label{r2}
\end{equation}
\noindent where $\omega_2$ is a constant. This reward encourages the agent to obtain small serving set sizes, matching the objective in (\ref{objG}). At the end of one episode, the agent finishes selecting serving APs for all UEs. We provide a final reward as the weighted sum rate in (\ref{objSumRate}) scaled by the fairness index in (\ref{objFair}),\looseness=-1
\begin{equation}
r_3=\omega_3 \Phi \sum_{k=1}^K S_k=w_3 \frac{\left(\sum_{k=1}^K S_k\right)^3}{K \sum_{k=1}^K S_k^2},
\label{r3}
\end{equation}
\noindent where $\omega_3$ is a constant representing the weight and $\Phi'$ is the Jain's fairness index of $S_k$. The cumulative reward at the end of each episode is thus given by
\begin{equation}
r_{cu} = \sum^K_{k=1} (r_1 |\mathcal{M}_k| + r_2) + r_3.
\label{rcu}
\end{equation} 

\subsubsection{UnifSrv-heu}
\label{myHEU}
We also propose the algorithm \mbox{\textit{UnifSrv-heu}} in Alg. \ref{partialFair} as a heuristic solution with low complexity to (\ref{eq:optimization}). Since the throughput performance is correlated with the SNR of AP - UE channels, we use the average SNR $\beta_{mk}$ to represent the quality of the channel between AP $m$ and UE $k$ \cite{aging}. Consequently, in Steps 1 to 6, each UE connects to the AP with the best channel quality, i.e., the best SNR, to form the initial serving set. In Steps 7 to 10, a threshold $\alpha$ is calculated based on Jain's fairness index $\Phi$ of $S_k$. Since $\alpha$ corresponds to the $\left\lceil ((1-\Phi)K)^{th} \right \rceil$ smallest transformed SINR $S_k$ among all UEs, it represents the boundary below which the \mbox{worst-served} $(1-\Phi)K$ UEs fall, i.e., the fraction that lacks fairness. UEs with $S_k<\alpha$ are thus considered to be not fairly served and require more serving APs to be allocated. Therefore, in Steps 11 to 17, these UEs are assigned the next AP with the best SNR to their serving set, as long as both the AP and UE do not exceed the constraints given by (\ref{cstAP}) and (\ref{cstUE}). We also stop adding APs to the serving set if the total SNR of the links between serving APs and UE $k$ reaches $\delta$\% of the total SNR when all APs serve UE $k$. In this case, \mbox{\textit{UnifSrv-heu}} satisfies the objective (\ref{objG}) by excluding the APs that contribute only a marginal improvement in serving set channel quality. By contrast, UEs with $S_k \geq \alpha$ are considered to have already good serving sets, and therefore do not need to be assigned the $m^\text{th}$ additional serving AP. When the selection is done for all UEs, $\alpha$ will be evaluated again for the selection of the next $(m+1)^\text{th}$ AP. In the next iteration, the UEs that are not assigned a serving AP in the last iteration are likely to have an SINR lower than $\alpha$, and are thus prioritized to select a new serving AP. The algorithm thus assigns the priority of UEs in selecting serving APs in each iteration solely based on their SINR. The algorithm satisfies the objective (\ref{objSumRate}) by always selecting the APs with the best SNR to serve, and (\ref{objFair}) by applying the fairness-based threshold $\alpha$. Importantly, our algorithm thus prioritizes the AP selection for the worst-served UEs to improve their throughput, and thus the uniformity of the throughput performance over the CF-mMIMO network.

\begin{algorithm}[!tb]
\caption{UnifSrv-heu} \label{partialFair}
\textbf{Input:} The SNR $\beta_{mk}$ between AP $m$ and UE $k$ for \mbox{$m=1,...,M$} and $k = 1,...,K$. The zero matrix $D[t] = \mathbf{0}$. The pilot sequence length $\tau_p$. The maximum serving set size $G_{max}$. The SNR percentage threshold $\delta$.
\begin{algorithmic}[1]

\For {$k = 1,...,K$}

\State Sort the APs in descending order based on $\beta_{mk} [t]$, $m=1,...,M$.

\State Define the AP index in the descending order as $\{O(1),O(2),...,O(M) \}$.

\State Set $D_{O(1)k} = 1$.

\State Calculate $S_k$ with (\ref{simpleSINR}).

\EndFor

\For {$m = 2,...,M$}

\State Calculate the Jain's fairness index $\Phi$ with (\ref{objFair}), replace $R_k$ with $S_k$.

\State Sort $S_k$ for all UEs in the ascending order.

\State Set the threshold $\alpha = S_{\left\lceil ((1-\Phi)K)^{th} \right \rceil}$.

\For {$k = 1,...,K$}

\If {$S_k < \alpha$ \& $W_m \leq \tau_p$ \& $G_k \leq G_{max}$ \& $\sum_{m=1}^M D_{mk} \beta_{mk} < \delta \sum_{m=1}^M \beta_{mk}$}

\State Set $D_{O(m)k} = 1$.

\EndIf

\State Calculate $S_k$ with (\ref{simpleSINR}).

\EndFor

\EndFor

\end{algorithmic}
\textbf{Output:} The dynamic cooperation matrix $D$.
\end{algorithm}

\subsection{Reference AP Selection Algorithms}
\label{APref}

We here briefly present four reference AP selection algorithms as the representative benchmarks for existing works detailed in Sec. \ref{related}, as follows. 

\textbf{PUC}: The \textit{pure UE-centric (PUC)} algorithm from \cite{EE} is representative of the algorithms from \cite{EE, DRLmaxRate, DRLmaxMinRate} that only focus on maximizing the per-UE throughput. The UEs are served by the APs with the best SNR until the total SNR of the serving AP set reaches $\delta$\% of the original CF-mMIMO case (i.e., when all APs serve each UE). \looseness=-1

\textbf{PUC-const}: The \textit{constrained pure UE-centric (PUC-const)} algorithm from \cite{MAcf} adds the AP capacity constraint based on the \textit{PUC} algorithm. The candidate serving AP is still chosen according to channel quality, i.e., SNR, and therefore this algorithm also only considers maximizing the throughput as an objective. If an AP reaches capacity, i.e., is already assigned to $\tau_p$ UEs, the algorithm compares the SNR between the new candidate UE and the worst SNR among the assigned UEs and replaces the connection if the candidate UE has a better SNR. 

\textbf{CUC}: The \textit{clustered UE-centric (CUC)} algorithm from \cite{cfsim} is chosen as representative of the algorithms from \cite{cfsim, GMM, DRLreduceConnect} that aim at limiting the serving AP set size. The area is divided into $N_c$ disjoint square clusters and on average $Q=M/N_c$ APs are assigned to each cluster according to their locations. The network first finds $E$ APs with the best SNR to a UE $k$, then the APs in the clusters these $E$ APs belong to form the serving set of UE $k$.

\textbf{PF-DRL}: The \textit{proportionally-fair DRL-based (PF-DRL)} algorithm from \cite{DRLmobi} aims at providing proportional fairness in data rates to all UEs over the network. It applies a DRL-based solution for a single-objective optimization problem, which share the same constraints as (\ref{cstAP})-(\ref{constrain}) and the objective given by $\max \limits_{D_{mk}} \; \sum_{k=1}^K \log (R_k)$. We note that this algorithm is originally designed for a cellular network, where the state space only needs to observe the channel quality and serving capacity of one serving AP $m'$, i.e., $s_k = \{\beta_{m'k}, W_{m'}\}$. In order to support multiple APs serving a given UE in CF-mMIMO, we generalize the state space design of \textit{PF-DRL} to observe all current serving APs as in \textit{UnifSrv-DRL}. The action space and DQN algorithm of \textit{PF-DRL} is the same as our \textit{UnifSrv-DRL} since both algorithms provide the same action of selecting available serving APs under the same DQN framework, making \mbox{\textit{PF-DRL}} also satisfy the constraints in \eqref{cstAP} and \eqref{cstUE}. However, the \textit{PF-DRL} algorithm only has a single final reward function according to the single objective at the end of one episode. Namely, for \textit{PF-DRL}, the intermediate rewards $r_1=r_2=0$ and the final reward $r_3 = \sum_{k=1}^K \log (R_k)$.  \looseness=-1

In our performance evaluation in Sec. \ref{results}, we also consider two boundary reference cases, representing the upper and lower bound of the CF-mMIMO network throughput, respectively: the \textit{original \mbox{CF-mMIMO}} case, where all APs in the network serve all UEs (i.e., $G_k = M, \forall k$), and the \textit{small cell} case, where only one AP with the best SNR serves a UE.

\section{Results}
\label{results}
We conduct a comprehensive performance evaluation of \mbox{CF-mMIMO} in realistic urban networks. In Sec. \ref{benchmark} we analyse the throughput performance of the prior benchmark AP selection algorithms given in \mbox{Sec. \ref{APref}}. In Sec. \ref{unifsrv} we then evaluate the performance of our \textit{UnifSrv} AP selection algorithms proposed in Sec. \ref{algo}. In Sec. \ref{AlgoAna} we present a detailed analysis of our \textit{UnifSrv} algorithms.

We conduct our performance evaluation of \mbox{CF-mMIMO} by numerically calculating the per-UE throughput via (\ref{rate}), in line with prior works \cite{EE, DRLmaxRate, DRLmaxMinRate,  cfsim, GMM, DRLreduceConnect, MAcf}. We simulate $K=50$ UE tracks all with a mobility period of \mbox{400 s} and compute the channel quality of all AP - UE links in the network to select each UE's serving AP set once per communication block, i.e., every \mbox{$\tau_c=200$ ms}. We thereby obtain a distribution of UE throughput over $50 \times (400/0.2)=100000$ Monte-Carlo realizations. For the considered AP selection algorithms, we set \mbox{$\delta=95$\%} for the \textit{PUC} and our \textit{UnifSrv-heu} algorithms.  For the \textit{CUC} algorithm, for a fair comparison, we set \mbox{$\{E,Q\}$} to obtain the same average serving AP set size as the average $G_k$ of \mbox{\textit{UnifSrv-heu}} via simulation as: \mbox{$\{E,Q\} = \{7, 20\}$} to achieve an average serving AP set size $G=40$ in the network of Seoul and \mbox{$\{E,Q\} = \{3, 17\}$} to achieve $G=25$ in Frankfurt. For our \mbox{\textit{UnifSrv}} algorithms and \mbox{\textit{PF-DRL}}, we set the following parameters: $W_{max}=\tau_p=10$, $G_{max}=100$ for Seoul and $G_{max}=70$ for Frankfurt, outage SNR $\beta_0 = -20$ dB, learning rate of $10^{-5}$, discount factor of 0.99, batch size of 128, buffer size of $10^6$, the total number of training episodes is 700, and the number of steps in each episode $T=KU_m$ with the number of steps per round $U_m=100$. The neural network consists of \mbox{$L=2$} hidden layers with $n=512$ neurons each. Furthermore, we set the weights of the reward function of the DQN algorithm for \textit{UnifSrv-DRL} as $\omega_1=1$, $\omega_2 = 10$, and $\omega_3 = 2000$, as they were found to perform best. \looseness=-1

\subsection{Throughput Performance Evaluation of CF-mMIMO Under Realistic Urban Propagation}
\label{benchmark}
\begin{figure}[!tb]
\centering
\includegraphics[width=1\columnwidth]{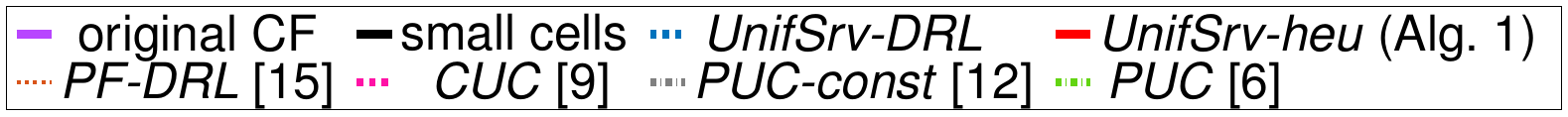}
	\subfigure[Seoul network (realistic raytracing channel model)]{
	    \label{SEseoul.seoul}
        \includegraphics[width=1\linewidth]{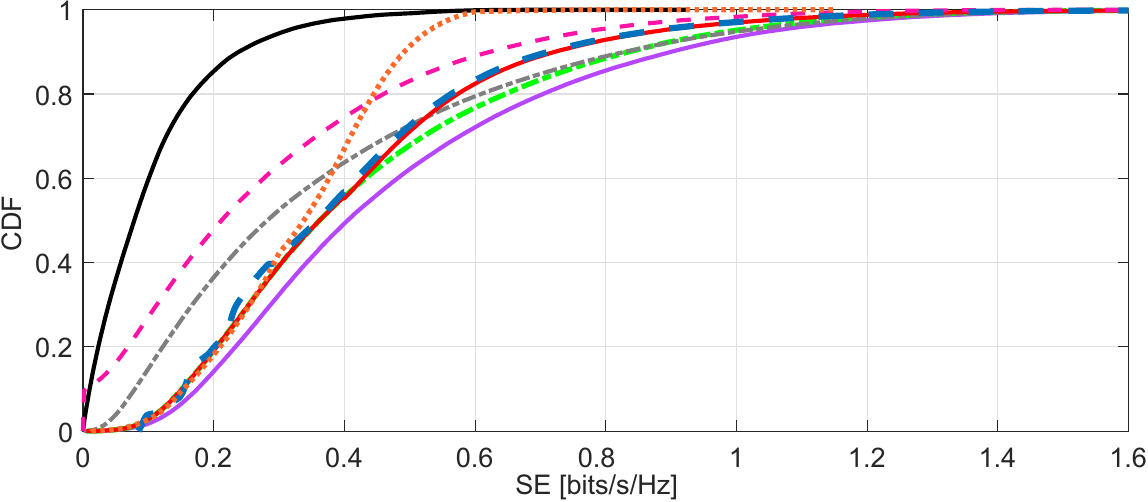}}
    \subfigure[PPP network (ideal log-distance channel model)]{
        \label{SEseoul.PPP}
        \includegraphics[width=1\linewidth]{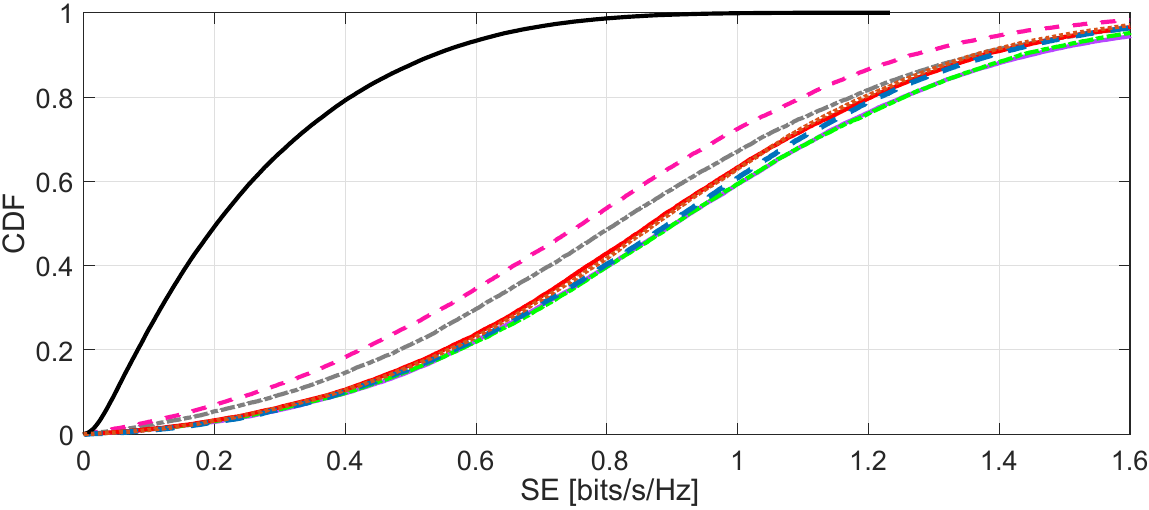}}
\caption{Spectral efficiency distribution over all UEs and their entire mobility periods with different AP selection algorithms for Seoul and PPP networks, with $M=324$ APs ($\lambda = 576 \text{ AP/km}^2$).}
\label{SEseoul}
\end{figure}

\begin{figure}[!tb] 
	\centering
	\subfigure[PPP, log-distance]{
		\label{snr.sub.1}
		\includegraphics[width=0.48\linewidth]{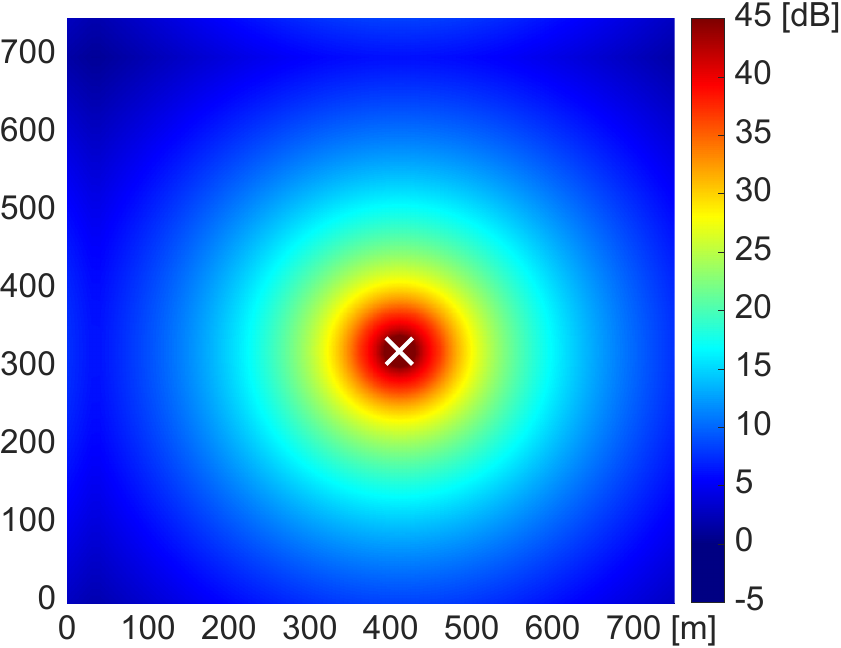}}
	\subfigure[Seoul, raytracing]{
		\label{snr.sub.2}
		\includegraphics[width=0.48\linewidth]{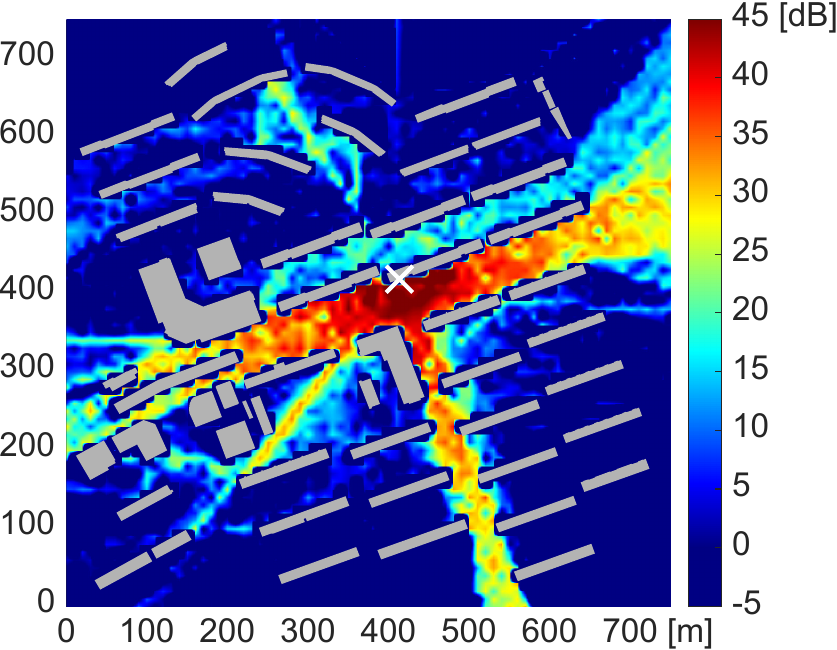}}
	\begin{minipage}[t]{0.48\linewidth}
		\subfigure[Frankfurt, raytracing]{
			\label{snr.sub.4}
			\includegraphics[width=\linewidth]{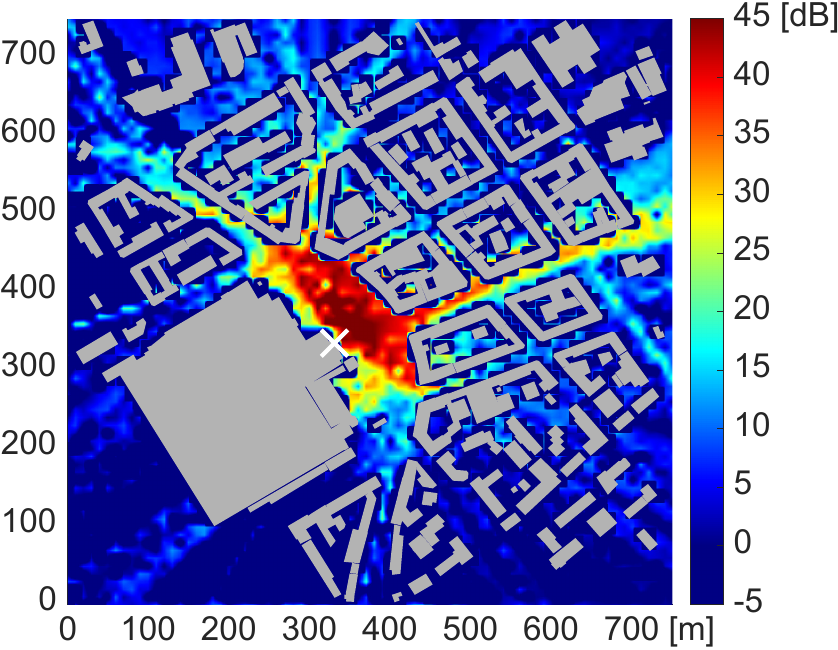}}
	\end{minipage}
	\hfill
	\begin{minipage}[t]{0.48\linewidth}
		\subfigure[SNR distribution over all APs]{
			\label{snrcdf}
			\includegraphics[width=0.87\linewidth]{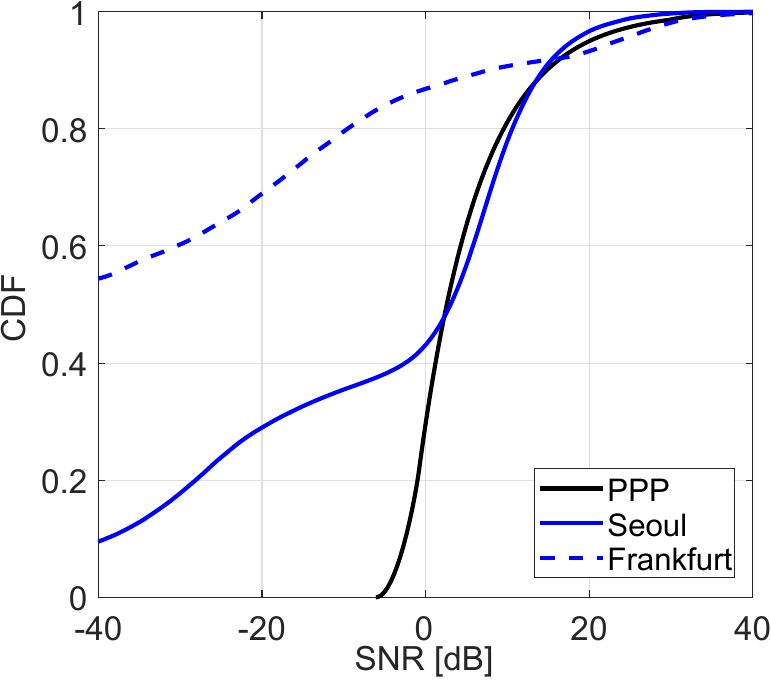}}
	\end{minipage}
	\caption{Spatial distribution of average SNR for different areas and channel models as a heatmap for an example AP (white cross) and CDF plots of all corresponding AP - UE channels. \looseness=-1}
	\label{snr}
\end{figure}

In this section, we study the throughput performance of \mbox{CF-mMIMO} under realistic versus ideal propagation, in order to establish the necessity of our problem formulation in \mbox{Sec.~\ref{problem}} and the \textit{UnifSrv} AP selection algorithm in Sec.~\ref{algo}. Fig. \ref{SEseoul} shows the distribution of the normalized throughput, i.e., spectral efficiency (SE) given by $R_k/B$, over all UEs and their entire mobility periods for all considered AP selection algorithms in the area of Seoul (Fig. \ref{SEseoul.seoul}) and its corresponding PPP network (Fig. \ref{SEseoul.PPP}). Let us first consider the performance of the two boundary cases: \mbox{Fig. \ref{SEseoul}} confirms that the \textit{original \mbox{CF-mMIMO}} case (where all APs serve all UEs) significantly improves the throughput compared to \textit{small cells}, for both the Seoul and PPP networks. In particular, \textit{original \mbox{CF-mMIMO}} significantly improves the throughput of the $95^{\text{th}}$ percentile (i.e., \mbox{worst-served}) UEs. Correspondingly, it achieves a fairness index of throughput per UE of $\Phi=0.8$ and $\Phi = 0.86$ in the Seoul and PPP networks, respectively, compared to $\Phi = 0.4$ and $\Phi = 0.5$ for \textit{small cells}. The original \mbox{CF-mMIMO} thus achieves both \textit{higher and more uniform} throughput than small cells, in both the Seoul and PPP networks. Moreover, comparing Fig. \ref{SEseoul.seoul} to Fig. \ref{SEseoul.PPP} shows that both architectures achieve significantly lower throughput in the urban case than in the PPP network. This throughput gap is caused by the channel quality difference illustrated in Fig. \ref{snr}. Different from the uniform spatial distribution of the average SNR in the PPP network with the log-distance path loss model shown in Fig. \ref{snr.sub.1}, \mbox{Figs. \ref{snr.sub.2}} and \ref{snr.sub.4} show that a large number of AP-UE links obtain very low SNR due to building shadowing in the urban areas. Namely, in realistic urban propagation, the channel quality is in general worse and more non-uniformly distributed than in PPP networks, as shown in Fig. \ref{snrcdf}. Consequently, the SE of both CF-mMIMO and small cells degrades in the case of realistic urban propagation. \looseness=-1

\begin{figure}[!tb]
\vspace{-0.2cm} 
	\centering
	\includegraphics[width=1\columnwidth]{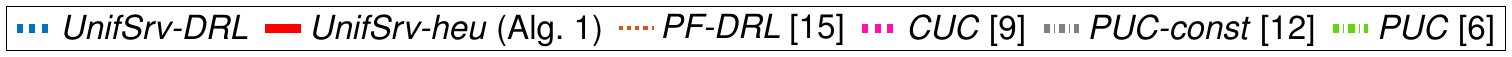}
	\subfigure[serving AP set size per UE, Seoul]{
		\label{serveSet.G}
		\includegraphics[width=0.48\linewidth]{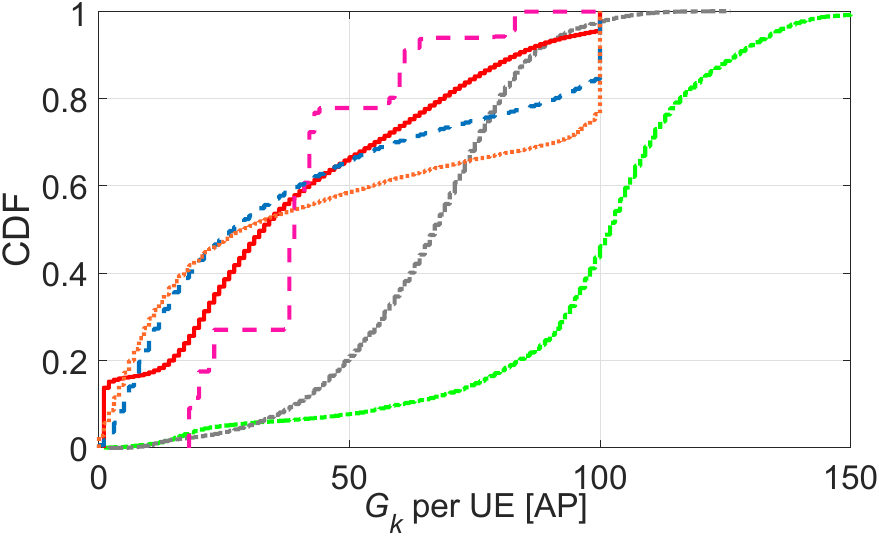}}
	\subfigure[served UEs per AP, Seoul]{
		\label{serveSet.W}
		\includegraphics[width=0.48\linewidth]{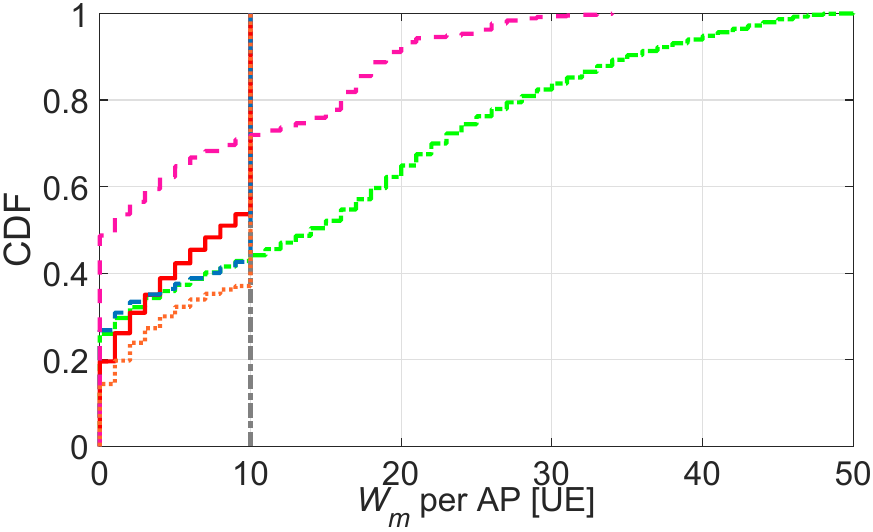}}
	\subfigure[serving AP set size per UE, PPP]{
		\label{serveSet.GPPP}
		\includegraphics[width=0.48\linewidth]{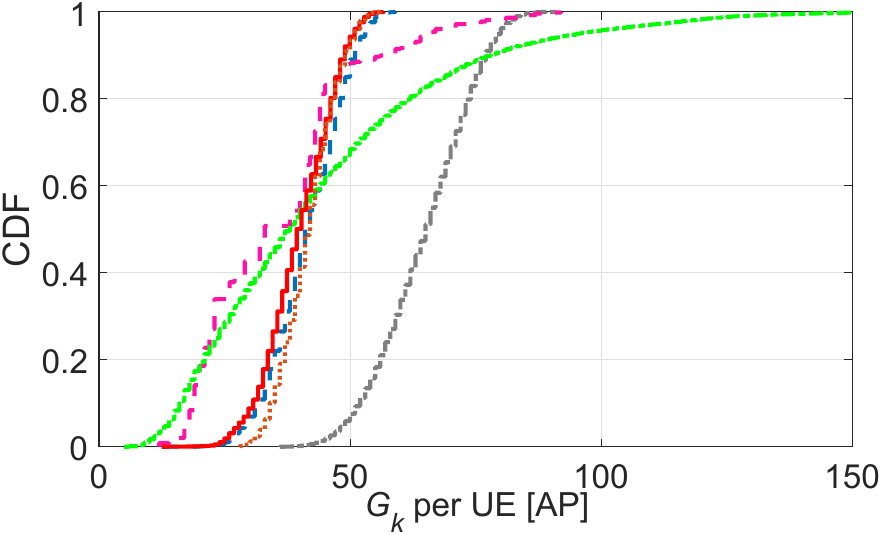}}
	\subfigure[served UEs per AP, PPP]{
		\label{serveSet.WPPP}
		\includegraphics[width=0.48\linewidth]{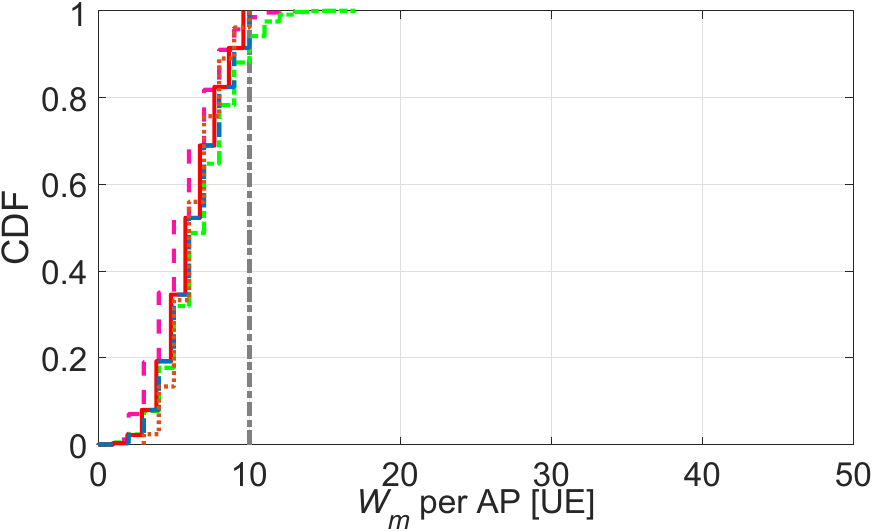}}
	\caption{Distribution of serving AP set size per UE and the number of served UEs per AP corresponding to the Seoul and PPP networks in Fig. \ref{SEseoul}.}
	\label{serveSet}
\end{figure}

\begin{figure}[!tb]
	\centering
	\subfigure[\textit{PUC}]{
		\label{scatter.minR}
		\includegraphics[width=0.46\linewidth]{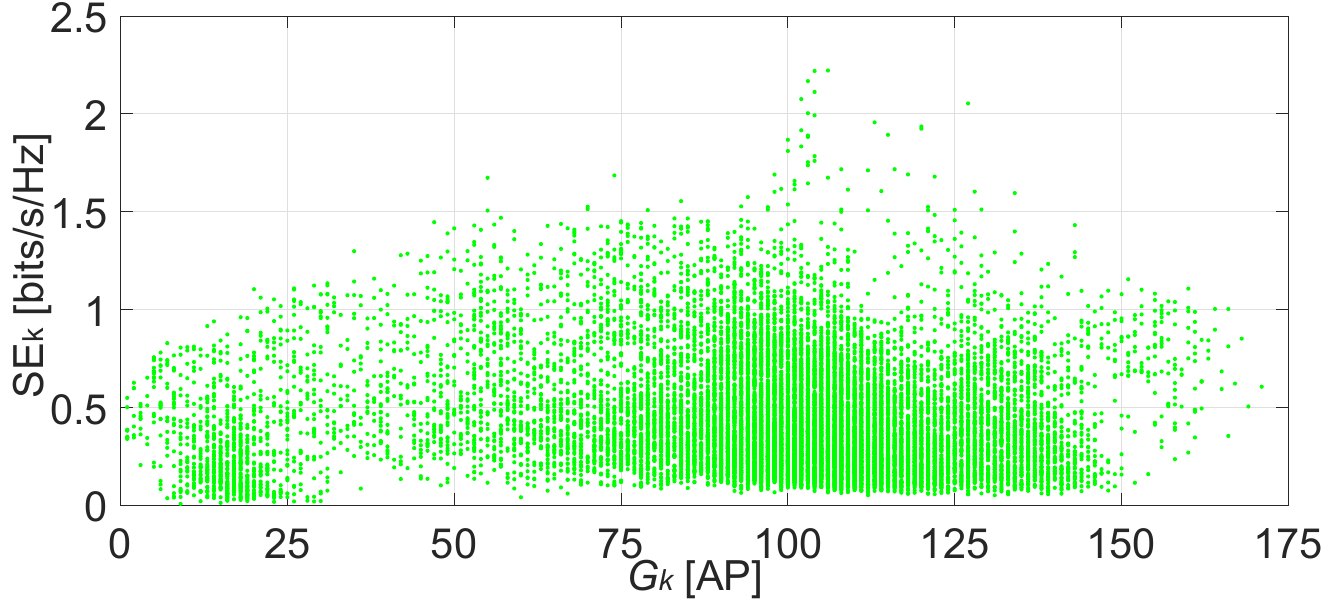}}
	\subfigure[\textit{PUC-const}]{
		\label{scatter.adaptive}
		\includegraphics[width=0.46\linewidth]{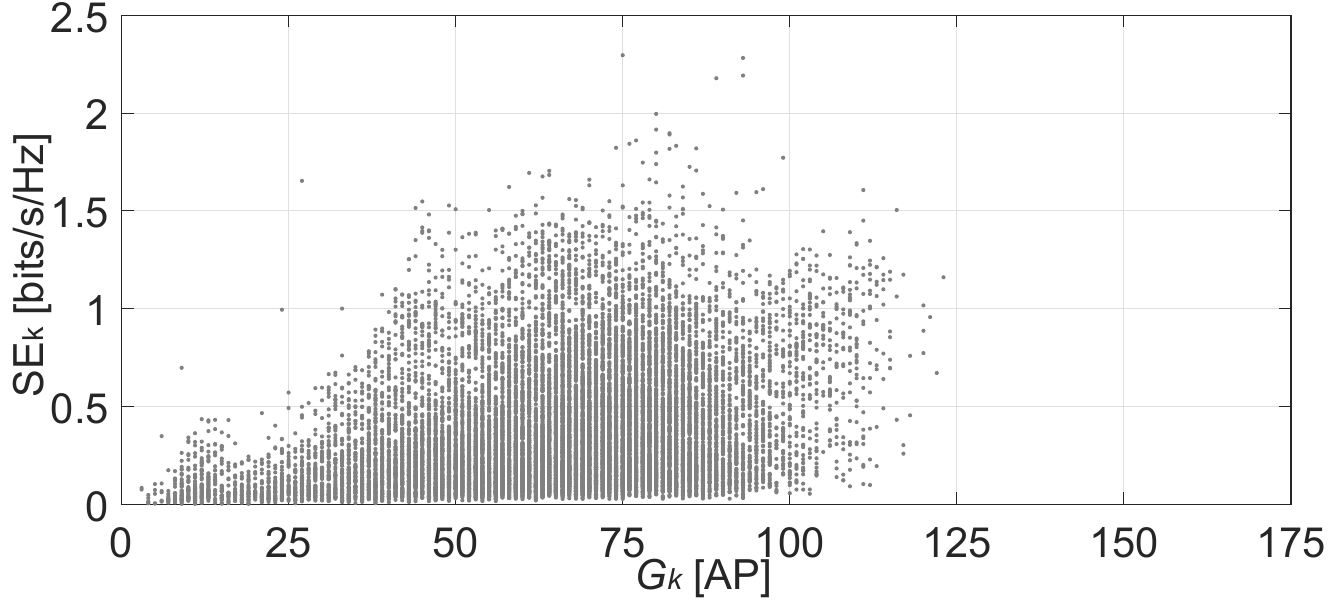}}
	\subfigure[\textit{PF-DRL}]{
		\label{scatter.PF}
		\includegraphics[width=0.485\linewidth]{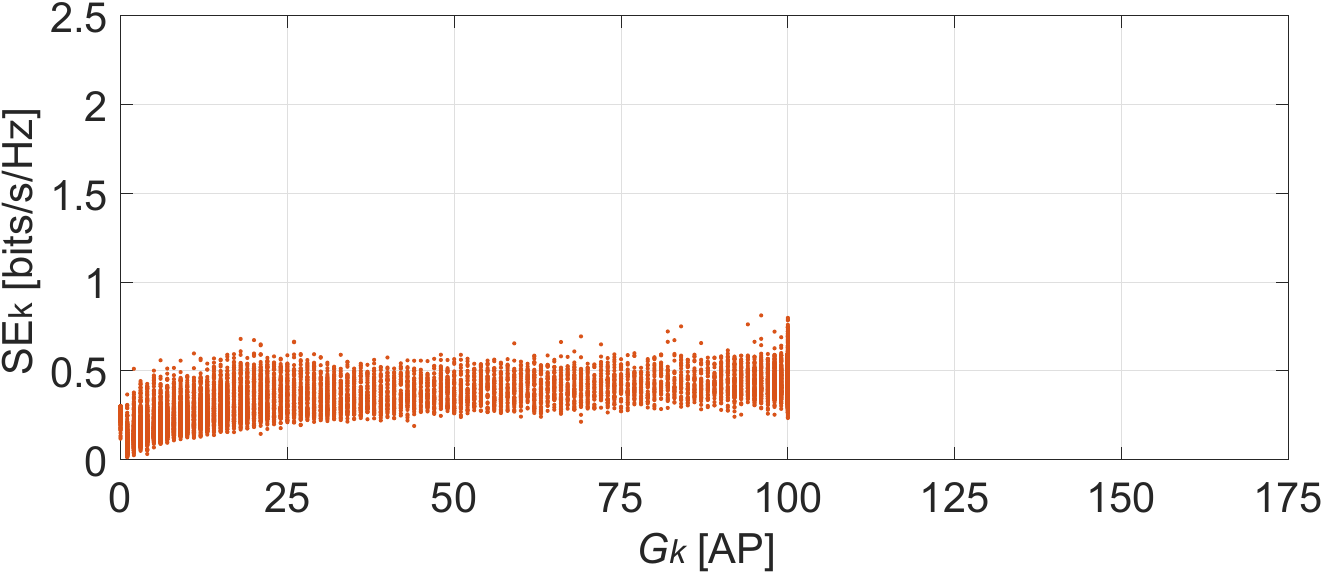}}
	\subfigure[\textit{CUC}]{
		\label{scatter.hybrid}
		\includegraphics[width=0.465\linewidth]{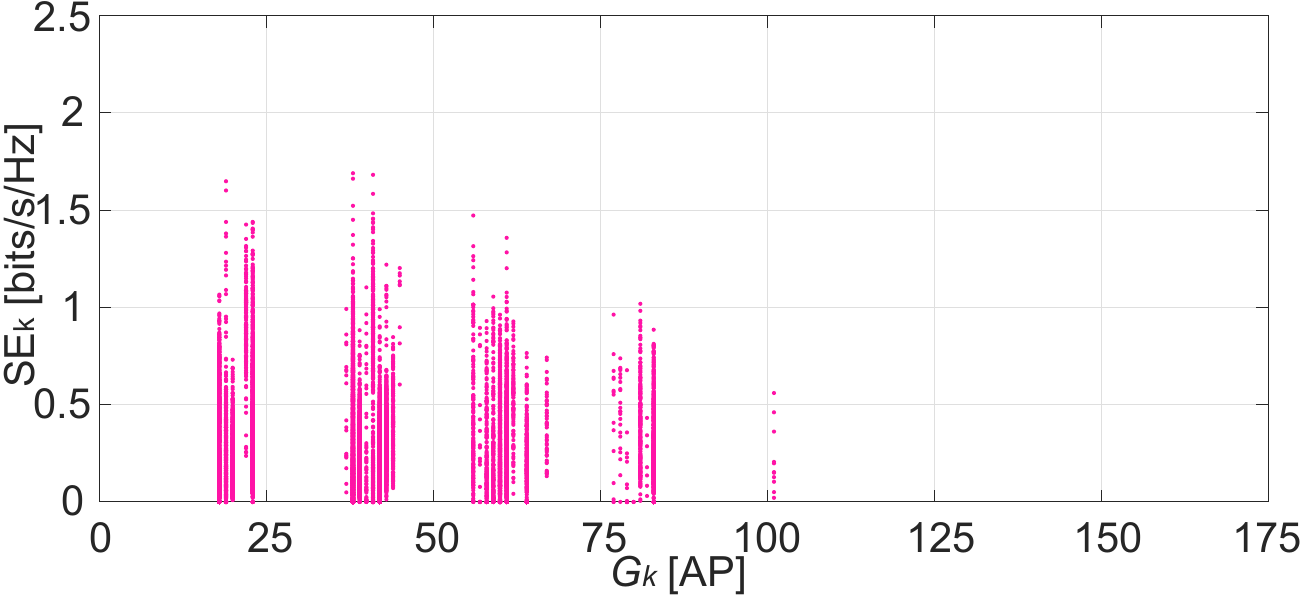}}
	\caption{Scatter plot of serving AP set size vs. spectral efficiency with different AP selection algorithms corresponding to the Seoul network in Fig. \ref{SEseoul.seoul}.\looseness=-1}
	\label{scatter}
\end{figure}

Let us next study the performance of the prior benchmark AP selection algorithms from Sec. \ref{APref}, with the SE shown in Fig. \ref{SEseoul} and the corresponding serving AP set sizes and numbers of served UEs per AP in Fig. \ref{serveSet}. To further illustrate the serving AP set selection mechanism of the considered algorithms in detail, we show in Fig. \ref{scatter} the scatter plot of serving AP set size per UE and its corresponding throughput. Fig. \ref{SEseoul.PPP} shows that in the PPP network, all benchmarks achieve a uniformly high throughput close to the \textit{original \mbox{CF-mMIMO}} upper bound and a significant improvement compared to \textit{small cells}, validating the performance of these AP selection schemes in the ideal PPP networks. Moreover, Figs.~\ref{serveSet.GPPP} and \ref{serveSet.WPPP} show in the PPP network, the constraints of $G_{max}=100$ and $W_{max}=10$ are only violated by \textit{PUC}, for a small proportion of UEs. The prior benchmarks thus are generally scalable in the PPP network.
 
Importantly, in the urban network by contrast, \mbox{Fig. \ref{SEseoul.seoul}} shows that among all considered prior benchmarks, only \textit{PUC} achieves a throughput close to the \textit{original \mbox{CF-mMIMO}} upper bound and similarly high throughput across UEs with different serving set sizes (\textit{cf. }\mbox{Fig. \ref{scatter.minR}}). However, this is at the cost of very large serving set size overall, violating both the constraints of serving set size $G_{max}=100$ and AP capacity $W_{max}=10$ for 60\% of UEs, as shown in \mbox{Figs. \ref{serveSet.G}} and \ref{serveSet.W}. The \textit{PUC} algorithm thus fails to select a scalable AP serving set in the realistic urban network. The \mbox{\textit{PUC-const}} algorithm limits the number of UEs served by an AP and reduces the serving AP set size compared to \textit{PUC}. However, this algorithm prioritizes serving the \mbox{well-performing} UEs by disconnecting the \mbox{poorly-performing} UEs. Consequently, \mbox{Fig. \ref{scatter.adaptive}} shows that with \mbox{\textit{PUC-const}}, only a large serving set size is correlated with achieving high throughput. As a result, in the urban network in \mbox{Fig. \ref{SEseoul.seoul}}, \mbox{\textit{PUC-const}} slightly improves the performance of the $10^\text{th}$ percentile (best-served) UEs compared to \textit{PUC}, but trades off half of the throughput of the $95^\text{th}$ percentile (worst-served) UEs to satisfy the constraint on AP capacity $W_m$, leading to a \mbox{non-uniform} throughput performance in the realistic urban network. Fig. \ref{SEseoul.seoul} shows that the \textit{PF-DRL} algorithm achieves a close throughput performance to \textit{PUC} for the 50\% worst-served UEs, but much worse throughput for the 50\% best-served UEs. This is because \textit{PF-DRL} only optimizes the proportional fairness of the throughput, where improving the data rate for the worst-served UEs is prioritized over achieving high sum data rate, so that it obtains low throughput for all serving set sizes (\textit{cf. }Fig. \ref{scatter.PF}). \textit{PF-DRL} thus leads to the opposite behavior to \textit{PUC-const}, i.e., improving the throughput of the \mbox{worst-served} UEs by trading off a reduction in throughput for the \mbox{best-served} UEs. Finally, \mbox{Fig. \ref{SEseoul.seoul}} shows that \textit{CUC} consistently yields poor throughput in the urban network. \mbox{Fig. \ref{scatter.hybrid}} further shows that with \textit{CUC}, a large serving set size is correlated with low throughput, due to the highly non-uniform channel conditions of urban propagation (\textit{cf.} Fig. \ref{snr.sub.2}), whereby the APs in the same cluster obtain very different channel conditions. Namely, even though the $E$ selected APs have good channel conditions, the serving CPU clusters these APs belong to do not provide overall good channel quality for the served UE \cite{mywcnc25}.

In summary, our results in Fig. \ref{SEseoul} reveal that although the prior benchmarks perform well in the ideal PPP network, in the realistic Seoul urban network, they either achieve poor throughput performance (\textit{PUC-const}, \mbox{\textit{PF-DRL}}, and \textit{CUC}) due to failing to assign appropriate APs to the UEs in need, or good throughput but at the cost of large serving set size due to assigning too many APs (\textit{PUC}). This emphasizes the need for a new AP selection approach that jointly optimizes the sum throughput, uniformity of throughput, and serving set size, in order to obtain the uniformly good performance of original \mbox{CF-mMIMO} in realistic urban networks, while staying scalable. \looseness=-1

\subsection{Performance of the Proposed \textit{UnifSrv} Algorithms}
\label{unifsrv}

Let us now evaluate the performance of our proposed \textit{UnifSrv} AP selection algorithms. Fig. \ref{SEseoul} shows that our \mbox{\textit{UnifSrv-DRL}} algorithm achieves very close throughput performance to the original \mbox{CF-mMIMO} upper bound for all UEs\footnote{This also shows that the transformation of SINR in (\ref{simpleSINR}) does not sacrifice optimality of our algorithms and is a valid optimization proxy.}, in both urban and PPP networks. Figs. \ref{serveSet.G} and \ref{serveSet.GPPP} show that our \textit{UnifSrv} algorithms obtain a similarly small serving AP set size both in the urban and PPP networks, in contrast to \textit{PUC}, which is the only considered benchmark in Fig. \ref{SEseoul} that achieves overall high throughput in the urban network but at the cost of large serving AP set size. Compared to the prior benchmarks \mbox{\textit{PUC-const}} and \textit{CUC} that partially satisfy the constraints on serving set size $G_k$ and AP capacity $W_m$, \mbox{\textit{UnifSrv-DRL}} achieves significantly higher throughput for the $95^{\text{th}}$ percentile (i.e., worst-served) UEs. Furthermore, compared to the benchmark \textit{PF-DRL}, which is the only considered scheme other than our \textit{UnifSrv} algorithms that satisfies both the constraints of \mbox{$G_k \leq G_{max}$} (Fig. \ref{serveSet.G}) and $W_m \leq W_{max}$ (\mbox{Fig. \ref{serveSet.W}}), \mbox{\textit{UnifSrv-DRL}} achieves significantly higher throughput for the 50\% best-served UEs, and therefore achieves higher sum throughput over the network. Fig. \ref{SEseoul.seoul} also shows that our \mbox{\textit{UnifSrv-heu}} heuristic algorithm achieves a comparable throughput performance to \mbox{\textit{UnifSrv-DRL}}, due to exhibiting a similar AP selection behavior, as illustrated in Figs. \ref{serveSet.G} and \ref{serveSet.W}. Therefore, unlike all prior benchmarks, both our \textit{UnifSrv} algorithms successfully achieve \textit{uniformly good} throughput performance close to the original CF-mMIMO upper bound in the urban network, with a minimized serving set size, ensuring scalability. \looseness=-1

\begin{figure}[!tb]
	\centering
    \subfigure[\textit{UnifSrv-DRL}]{
		\label{scatter.DQN}
		\includegraphics[width=0.485\linewidth]{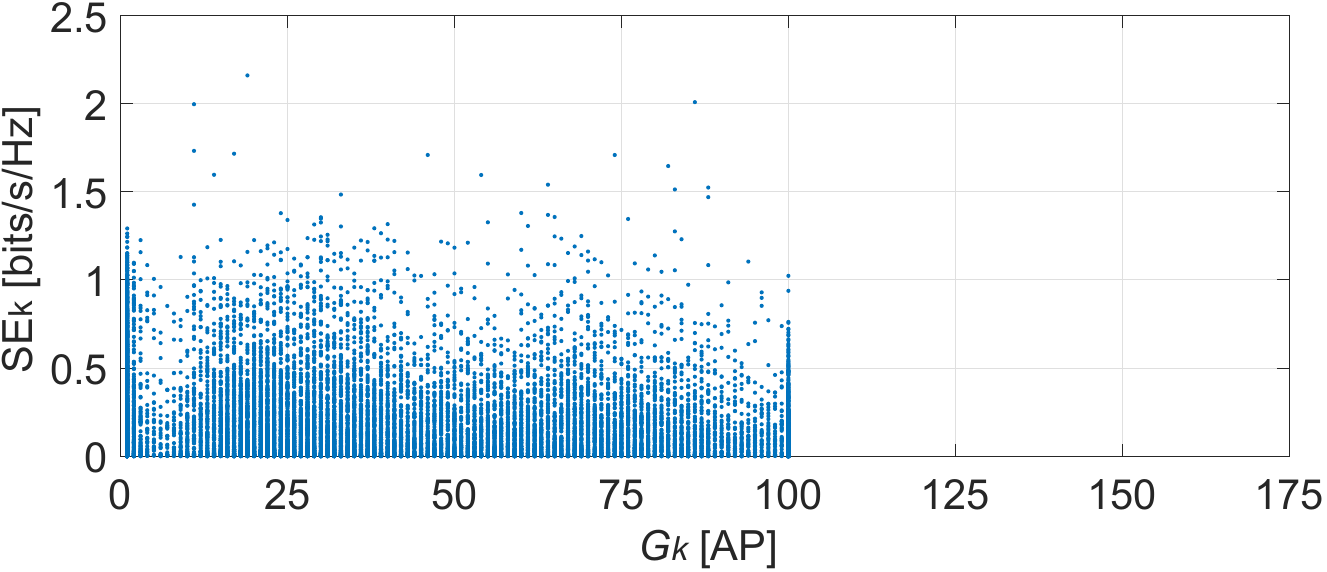}}
	\subfigure[\textit{UnifSrv-heu} (Alg. \ref{partialFair})]{
		\label{scatter.partialFair}
		\includegraphics[width=0.46\linewidth]{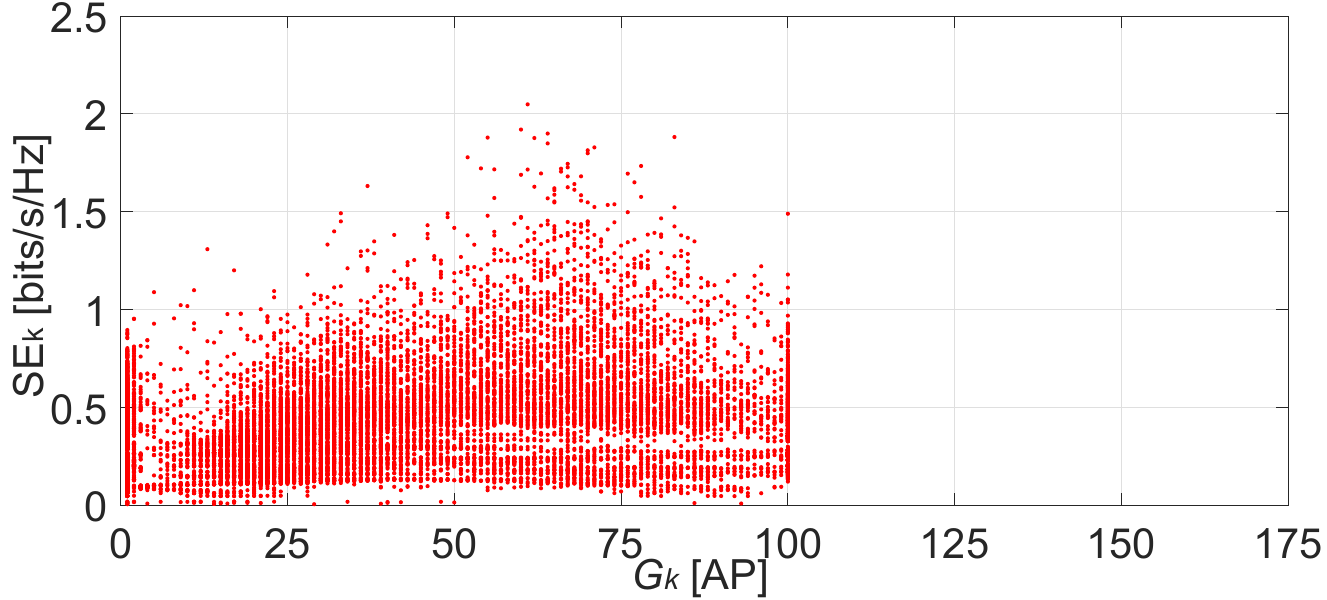}}
	\caption{Scatter plot of serving AP set size versus spectral efficiency with \textit{UnifSrv} corresponding to the networks in Fig. \ref{SEseoul.seoul}.}
	\label{scatterDQN}
\end{figure}

\begin{figure}[!tb]
	\centering
	\includegraphics[width=1\linewidth]{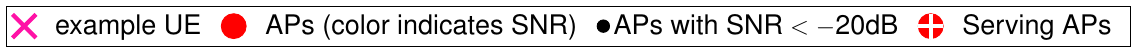}
	\subfigure[\textit{UnifSrv}, 95\%-ile UE]{
		\label{serveHeatDQN.bad}
		\includegraphics[width=0.48\linewidth]{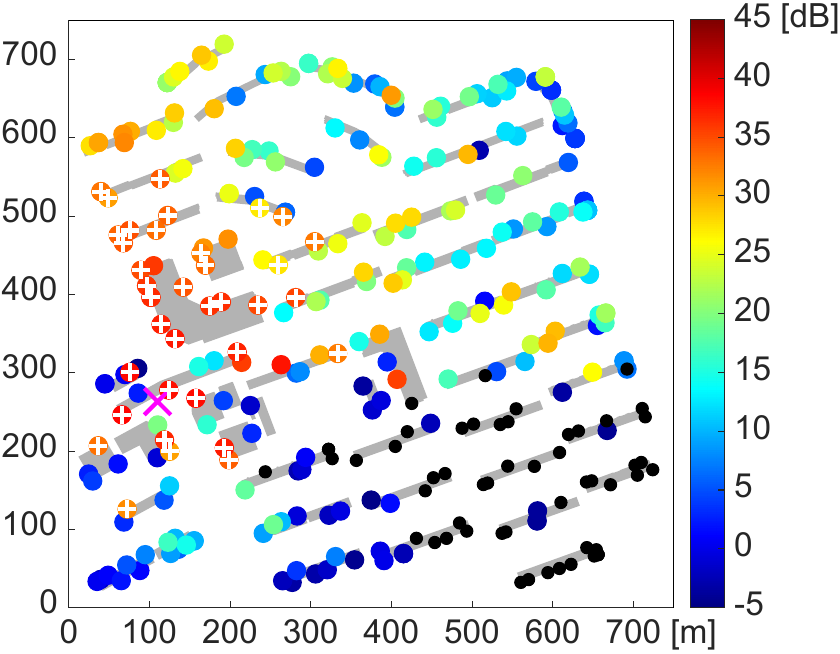}}
	\subfigure[\textit{UnifSrv}, 10\%-ile UE]{
		\label{serveHeatDQN.good}
		\includegraphics[width=0.48\linewidth]{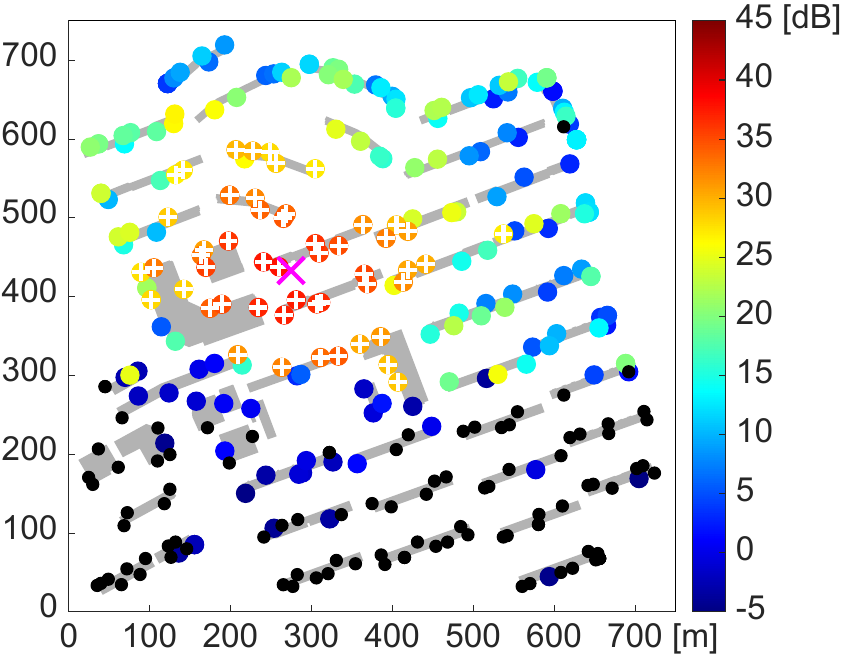}}
	\subfigure[\textit{PF-DRL}, 95\%-ile UE]{
		\label{serveHeatPF.bad}
		\includegraphics[width=0.48\linewidth]{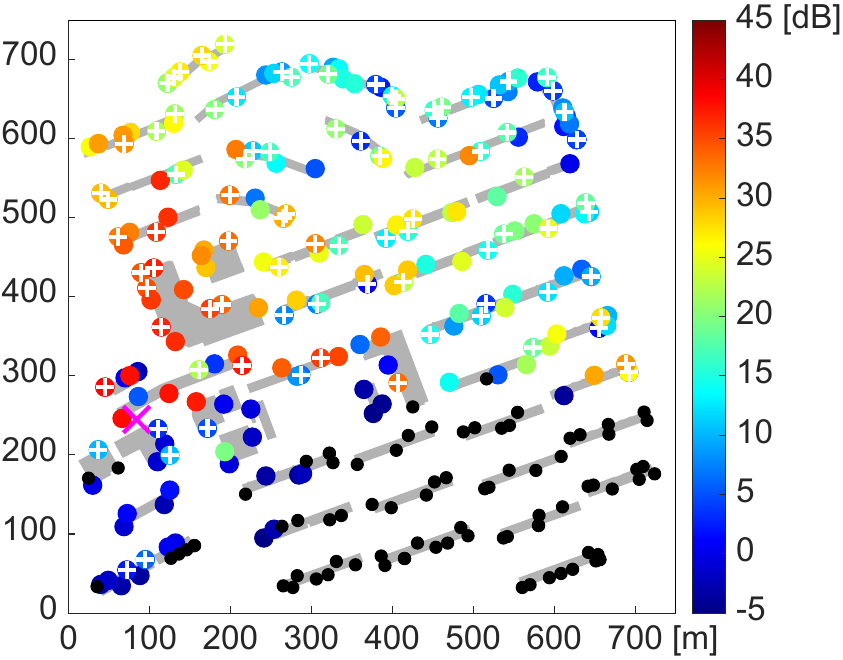}}
	\subfigure[\textit{PF-DRL}, 10\%-ile UE]{
		\label{serveHeatPF.good}
		\includegraphics[width=0.48\linewidth]{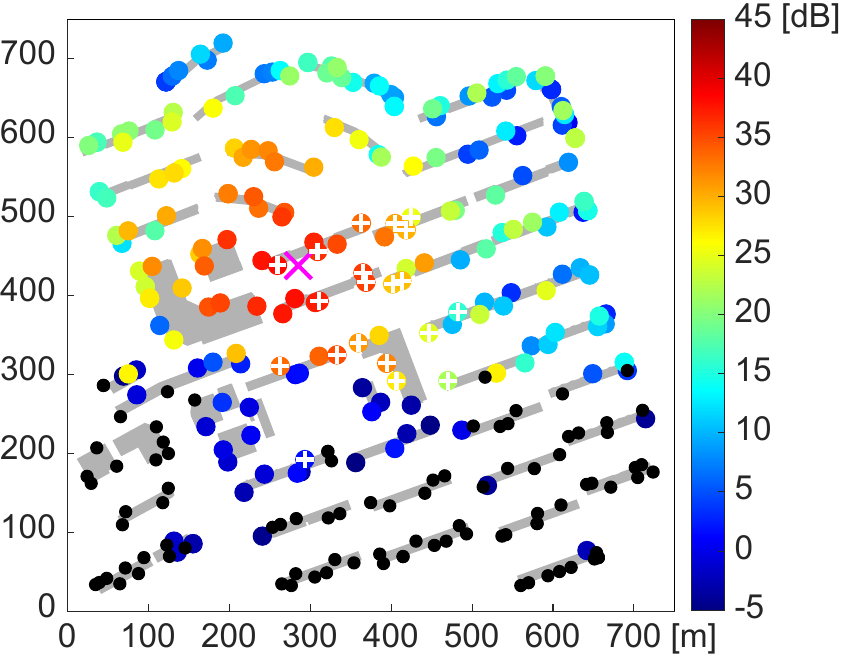}}
	\caption{Serving AP set SNR for an example 95\%-ile UE and 10\%-ile UE in Fig. \ref{SEseoul.seoul}, with (a) - (b) \textit{UnifSrv-DRL} (\textit{UnifSrv-heu} gives similar results) and (c) - (d) the \textit{PF-DRL} benchmark.\looseness=-1}
	\label{serveHeat}
\end{figure}

\looseness=-1

We further illustrate the AP selection of \textit{UnifSrv} with the scatter plot in \mbox{Fig. \ref{scatterDQN}} and the SNR heatmap in Fig. \ref{serveHeat}. Both our \textit{UnifSrv-DRL} algorithm in \mbox{Fig. \ref{scatter.DQN}} and our \textit{UnifSrv-heu} algorithm in Fig. \ref{scatter.partialFair} achieve similar throughput for different serving set sizes that do not violate the constraint of $G_{max}$. Figs. \ref{serveHeatDQN.bad} and \ref{serveHeatDQN.good} illustrates that they do so by successfully assigning a small serving AP set with near-optimal channel conditions to serve different UEs. In particular, \mbox{\textit{UnifSrv-DRL}} selects the optimal serving set by maximizing the reward function in (\ref{rcu}), which maximizes both the total data rate and fairness index and minimizes the serving set size, while \mbox{\textit{UnifSrv-heu}} does so by introducing a fairness \mbox{index-based} threshold to control the serving AP set expansion. By contrast, Figs. \ref{serveHeatPF.bad} and \ref{serveHeatPF.good} show that \textit{PF-DRL} -- the only benchmark that satisfies the $G_{max}$ and $W_{max}$ constraints -- assigns a very large serving set size to the worst-served UE for good channel conditions of the serving set, but a very small serving set size to the best-served UEs to satisfy the constraints, thereby trading off sum throughput for fairness. We emphasize that in contrast to the prior benchmarks, our \textit{UnifSrv} algorithms achieve their superior performance by \textit{jointly} maximizing both the throughput via \eqref{objSumRate} and the fairness of the throughput via \eqref{objFair}, while achieving scalability by minimizing the serving set size via \eqref{objG}. As a result, our \textit{UnifSrv} algorithms significantly improve the performance of the worst-served UEs, while only trading off a very small amount of data rate (throughput difference with \textit{PUC} being less than 0.1 bits/s/Hz in Fig. \ref{SEseoul.seoul}) of the best-served UEs to satisfy the $G_{max}$ constraint and obtain a small serving set size of on average 25 APs. \mbox{Fig. \ref{serveSet.G}} further shows that compared to the PPP network in \mbox{Fig. \ref{serveSet.GPPP}}, our \textit{UnifSrv} algorithms in the urban network result in a wider distribution of serving set size per UE with the same average $G$ value. This illustrates that our algorithms achieve their uniformly good throughput performance in the non-uniform urban environment by assigning the serving APs with both the appropriate quantity and quality to each UE based on channel conditions and system cost constraints.

\begin{figure}[!tb]
\centering
\includegraphics[width=1\columnwidth]{legendSE.pdf}
\includegraphics[width=1\linewidth]{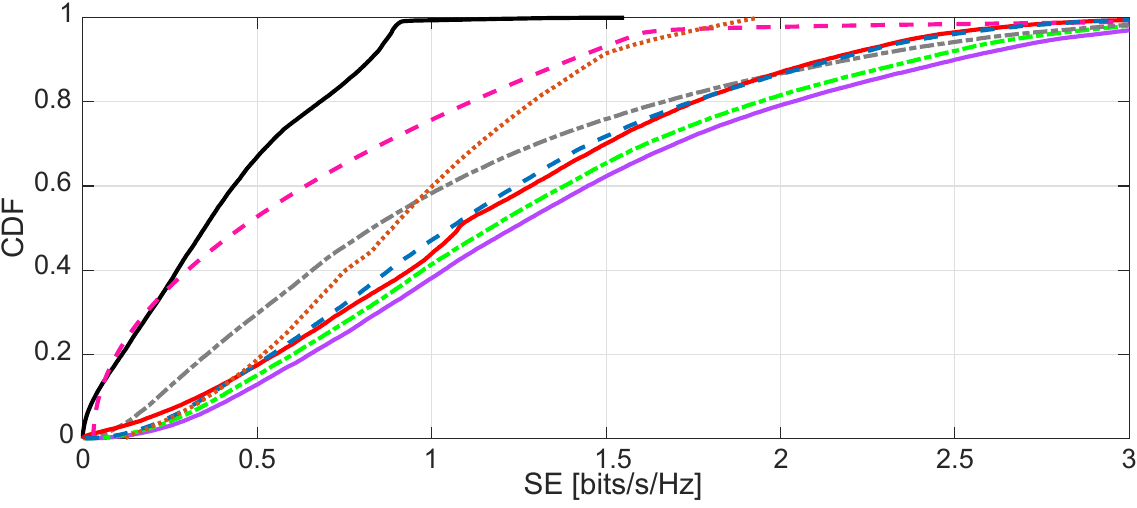}
\caption{Spectral efficiency distribution over all UEs and their entire mobility periods with different AP selection algorithms for the Frankfurt network, with $M=625$ APs ($\lambda = 1111 \text{ AP/km}^2$).}
\label{SEfrankfurt}
\end{figure}

\begin{figure}[!tb]
	\centering
	\includegraphics[width=1\columnwidth]{legend1.pdf}
	\subfigure[serving AP set size per UE]{
		\label{serveSetFL.G}
		\includegraphics[width=0.48\linewidth]{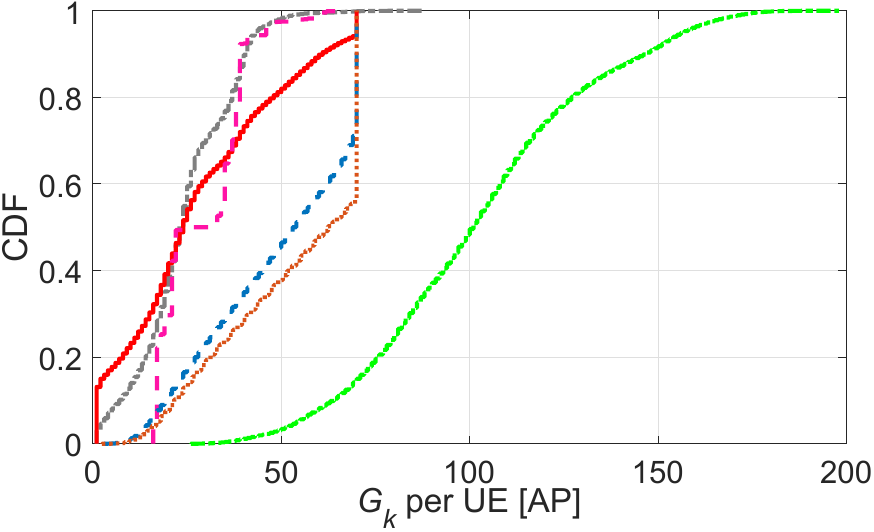}}
	\subfigure[served UEs per AP]{
		\label{serveSetFL.W}
		\includegraphics[width=0.48\linewidth]{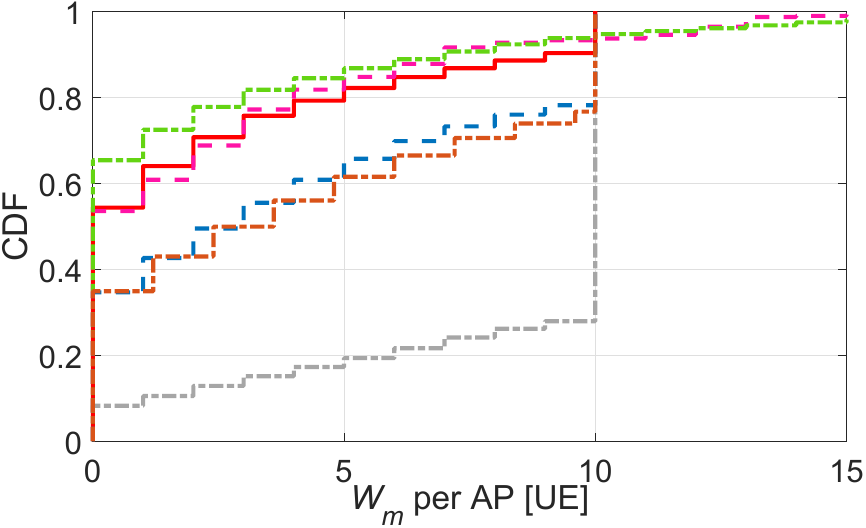}}
	\caption{The distribution of serving AP set size per UE and the number of served UEs per AP corresponding to the Frankfurt networks in Fig. \ref{SEfrankfurt}.}
	\label{serveSetFL}
\end{figure}

Finally, we study the performance of CF-mMIMO in Frankfurt (\textit{cf.} Figs. \ref{topo.frankfurt} and \ref{snr.sub.4}), which represents an even more \mbox{non-uniform} urban propagation environment than Seoul, with narrower streets and more pronounced shadowing and urban canyon effects. \mbox{Fig. \ref{SEfrankfurt}} shows that the throughput performance trends in Frankfurt are similar to those for Seoul in \mbox{Fig. \ref{SEseoul.seoul}} and that our \textit{UnifSrv} algorithms also achieve throughput close to the original \mbox{CF-mMIMO} upper bound. Fig. \ref{SEfrankfurt} shows that in Frankfurt, the \textit{CUC} algorithm performs worse than in Seoul, resulting in \mbox{CF-mMIMO} performing \textit{worse} than small cells for around 40\% of UEs. Furthermore, with the \textit{PF-DRL} algorithm, more UEs (around 80\%) obtain worse throughput than \textit{UnifSrv} and \textit{PUC}, although \mbox{Fig. \ref{serveSetFL}} shows that \textit{PF-DRL} selects similarly large serving set sizes as \textit{UnifSrv-DRL}. This indicates that \mbox{\textit{PF-DRL}} tends to assign more APs to serve the worst-served UEs in order to compensate the overall poorer channel conditions in Frankfurt (\textit{cf.} \mbox{Fig. \ref{snrcdf}}), while leaving more UEs with insufficient serving sets. This emphasizes that it is not sufficient to solely optimize the proportional fairness of the throughput as \mbox{\textit{PF-DRL}} does. Importantly, \mbox{Figs. \ref{SEfrankfurt}} and \ref{serveSetFL} show that both our \mbox{\textit{UnifSrv-DRL}} and \mbox{\textit{UnifSrv-heu}} achieve comparable throughput while satisfying the constraints of $G_{max}$ and $W_{max}$ also in Frankfurt. Fig. \ref{serveSetFL.G} shows that the DRL solution \mbox{\textit{UnifSrv-DRL}} tends to choose a slightly larger serving set size compared to the heuristic \textit{UnifSrv-heu}. This is because in Frankfurt, the channel quality is in general very poor (\textit{cf.} Fig. \ref{snrcdf}), so that the agent requires more serving APs per UE to achieve a high reward of $r_1$. We note that due to this slightly larger serving set size, \mbox{\textit{UnifSrv-DRL}} achieves slightly better throughput for the $15^\text{th}$ percentile UEs than \textit{UnifSrv-heu} in Frankfurt. \looseness=-1

\subsection{Analysis of the Proposed \textit{UnifSrv} Algorithms}
\label{AlgoAna}

Having established the superior throughput performance of \textit{UnifSrv} in realistic urban networks in Sec. \ref{unifsrv}, we further analyse our proposed algorithms in terms of convergence, feasibility, optimality over multiple objectives, generalization ability in different networks, and system cost.

\subsubsection{Convergence}

\begin{figure}[!tb]
	\centering
	\subfigure[Seoul]{
		\label{converge.SL}
		\includegraphics[width=0.48\linewidth]{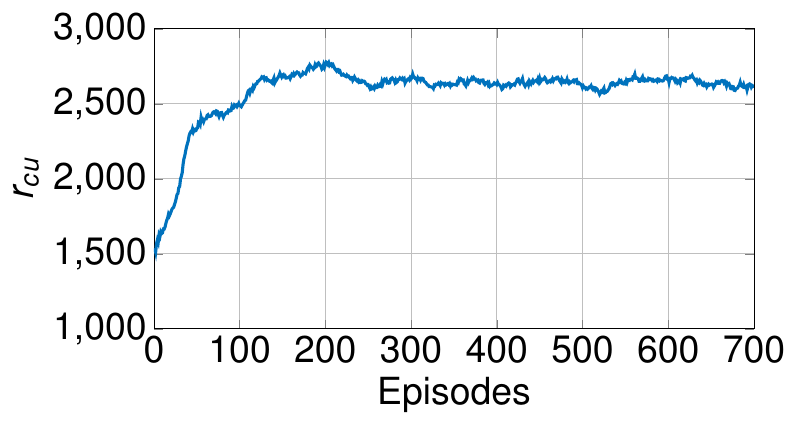}}
	\subfigure[Frankfurt]{
		\label{converge.FL}
		\includegraphics[width=0.48\linewidth]{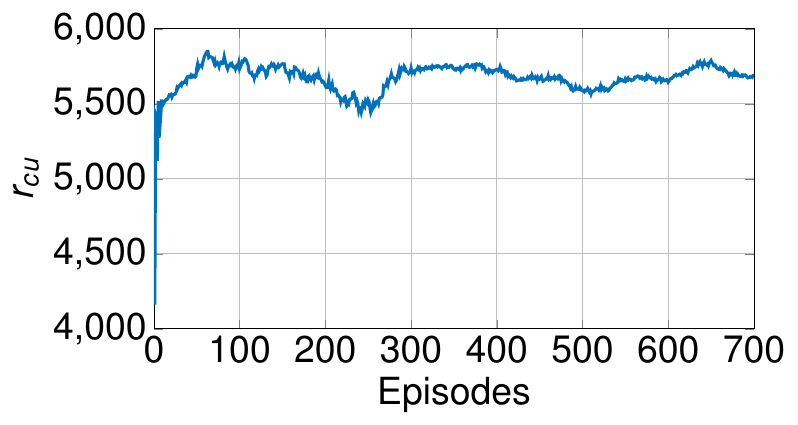}}
	\caption{Transition of the cumulative reward in training for \textit{UnifDrv-DRL} in Seoul and Frankfurt networks.}
	\label{converge}
\end{figure}

Our heuristic algorithm \textit{UnifSrv-heu} contains two for-loops with finite lengths and is guaranteed to terminate after $K+(M-1)K$ loops (\textit{cf. }\mbox{Alg. \ref{partialFair}}) and output a feasible cooperation matrix $D$. We analyse the convergence of our \mbox{DRL-based} algorithm \textit{UnifSrv-DRL} as follows. Since theoretical convergence guarantees for DQN remain an open problem, we instead empirically assess the convergence of \mbox{\textit{UnifSrv-DRL}} by monitoring the cumulative reward $r_{cu}$ obtained at the end of each training episode. \mbox{Fig. \ref{converge}} shows the transition of $r_{cu}$ in the networks of Seoul and Frankfurt over the total 700 training episodes. Fig.\ref{converge.SL} shows that in the Seoul network, $r_{cu}$ increases rapidly over episodes \mbox{0 - 150} from random exploration, the rate of change slows down over episodes 150 -- 250 as the DQN fine-tunes its behavior, and finally converges after episode 250 remaining within $\pm 1.7\%$ of its mean for the remainder of training. A similar trend of $r_{cu}$ is observed for the Frankfurt network in Fig.~\ref{converge.FL}: rapid increase over episodes \mbox{0 - 50}, refinement over episodes \mbox{50 - 300}, and convergence after episode~300 with a fluctuation of $\pm 1.5\%$ of its mean for the remainder of training. \mbox{Fig. \ref{converge}} thus demonstrates the strong convergence behavior of \mbox{\textit{UnifSrv-DRL}}. \looseness=-1

\subsubsection{Feasibility \& Optimality}

\begin{figure}[!tb]
	\centering
	\subfigure[sum data rate]{
	     \label{objective.rate}
	     \includegraphics[width=1\columnwidth]{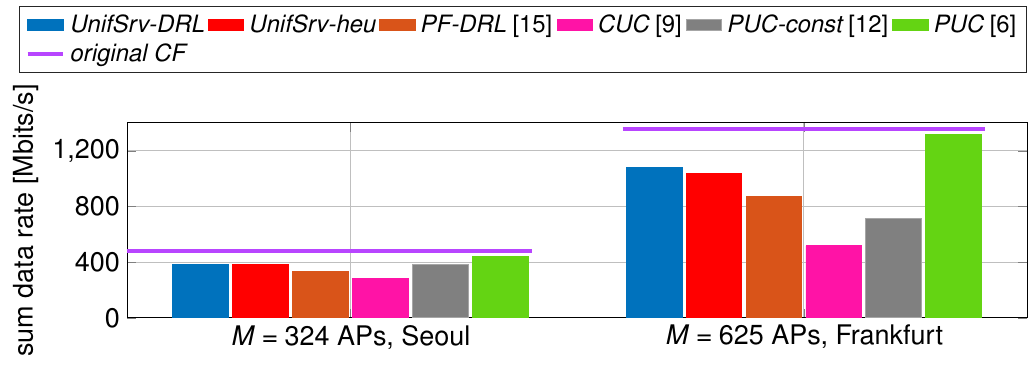}}
    \subfigure[fairness index $\Phi'$]{
	     \label{objective.fair}
	     \includegraphics[width=1\columnwidth]{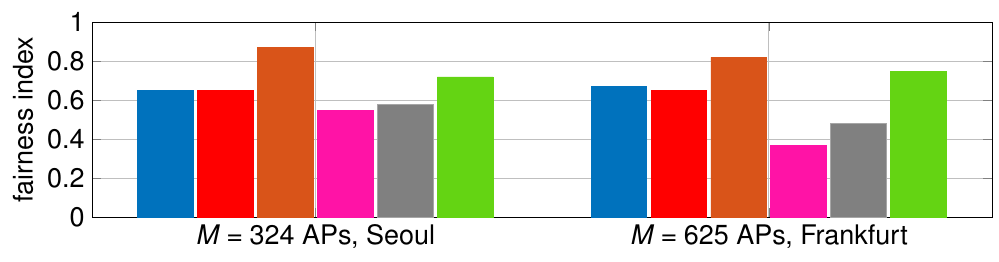}}	     
	\subfigure[average serving AP set size]{
	     \label{objective.G}
	     \includegraphics[width=1\columnwidth]{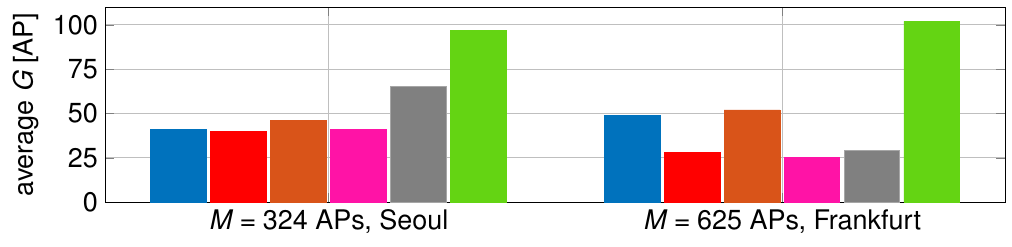}}
	\caption{The objective values of all algorithms in Seoul and Frankfurt.}
	\label{objective}
\end{figure}

\begin{table}[!tb]
\centering
\caption{AP Selection Constraints for Different Algorithms}
\setlength{\tabcolsep}{6pt}

\begin{tabular}{|p{2cm}|c|c|}
\hline
\textbf{Algorithms} 
& \makecell{\textbf{Serving Set Constraint}\\$G_{\max}$ \textbf{Met?}} 
& \makecell{\textbf{AP Capacity Constraint}\\$W_{\max}$ \textbf{Met?}}
\\ \hline

\textit{PUC} \cite{EE} 
& No 
& No 
\\ \hline

\textit{PUC-const} \cite{MAcf} 
& No 
& Yes 
\\ \hline

\textit{PF-DRL} \cite{DRLmobi}
& Yes 
& Yes 
\\ \hline

\textit{CUC} \cite{cfsim}
& Yes 
& No 
\\ \hline

\textbf{\textit{UnifSrv-DRL}}
& Yes 
& Yes 
\\ \hline

\textbf{\textit{UnifSrv-heu}}
& Yes 
& Yes 
\\ \hline

\end{tabular}
\label{perSum}
\end{table}

We here analysize the feasibility and optimality of our \textit{UnifSrv} algorithms over the multiple objectives shown in (\ref{eq:optimization}). Table \ref{perSum} shows that among all considered AP selection algorithms, only our \textit{UnifSrv} algorithms and \mbox{\textit{PF-DRL}} are feasible in terms of no constraint violation. However, Fig. \ref{objective} shows that for all three objectives in (\ref{eq:optimization}), the \mbox{\textit{PF-DRL}} benchmark only achieves a higher fairness index than \textit{UnifSrv}, i.e., performs well on objective \eqref{objFair}, due to solely maximizing the proportional fairness (\textit{cf. }Fig.~\ref{objective.fair}). It obtains significantly lower sum throughput (\textit{cf. }Fig.~\ref{objective.rate}) and higher serving set size (\textit{cf. }Fig. \ref{objective.G}) than \textit{UnifSrv}, demonstrating its poor performance on objectives \eqref{objSumRate} and \eqref{objG}. To further illustrate the optimality of \textit{UnifSrv}, we also in Fig. \ref{objective} benchmark it against the other algorithms that are infeasible with respect to the constraints in \eqref{eq:optimization} but achieve good performance on individual objectives. Fig. \ref{objective.rate} shows that both our \textit{UnifSrv-DRL} and \mbox{\textit{UnifSrv-heu}} algorithms achieve around 80\% of the sum data rate of the \textit{original \mbox{CF-mMIMO}} case, which is the upper bound of objective \eqref{objSumRate}, in both the Seoul and Frankfurt networks. Only the \textit{PUC} benchmark achieves slightly higher sum data rate than \textit{UnifSrv} since it is designed to solely maximize throughput. However, this is at the cost of a very large serving set size (\textit{cf. }Fig. \ref{objective.G}) and the violation of both constraints (\ref{cstAP}) and (\ref{cstUE}) (\textit{cf. }Table~\ref{perSum}). \mbox{Fig. \ref{objective.G}} shows that both \textit{UnifSrv} algorithms obtain low average serving AP set size $G$, the value of objective (\ref{objG}) divided by the number of UEs $K=50$, in both cities, confirming \textit{UnifSrv}'s good performance on limiting the \mbox{AP - UE} connections. The \textit{CUC} benchmark obtains the lowest $G$ and performs the best for objective \eqref{objG}. However, it obtains both low throughput (\textit{cf. }Fig. \ref{objective.rate}) and fairness index (\textit{cf. }Fig. \ref{objective.fair}), and thus performs poorly in the other two objectives in \eqref{eq:optimization}. It is also infeasible due to the violation of the constraint \eqref{cstAP} (\textit{cf. }Table \ref{perSum}). Finally, the \textit{PUC-const} benchmark performs poorly in all three objectives and is infeasible due to the violation of constraint \eqref{cstUE}. Overall, all benchmarks perform poorly in either throughput or scalability, due to the incomplete optimization objectives in their design, whereas \textit{UnifSrv} provides the first solution that balances all objectives, i.e., high throughput, high fairness index, and low serving set size, while incurring no constraint violation and guaranteeing feasibility.

\subsubsection{Generalization Ability}
\label{general}

\begin{figure}[!tb]
	\centering
	\includegraphics[width=1\columnwidth]{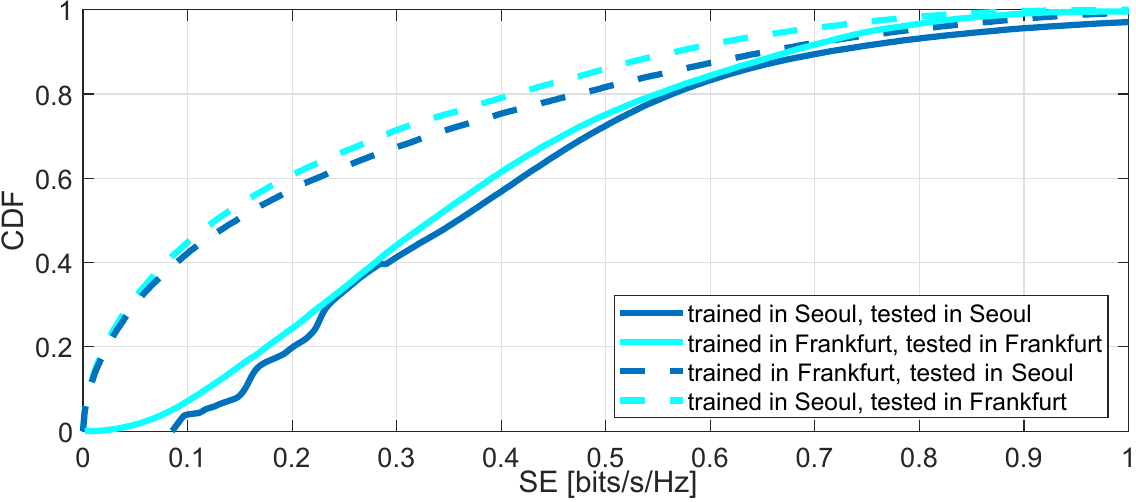}
	\caption{Throughput performance of agents of \textit{UnifSrv-DRL} trained and performs in Seoul and PPP networks, with $M=324$ APs ($\lambda = 576 \text{ AP/km}^2$).}
	\label{DRLcompare}
\end{figure}

In our \textit{UnifSrv-DRL} algorithm, the state and action spaces are designed for a specific network deployment and scale with the number of APs and UEs. Namely, a trained DRL agent for a given network cannot provide feasible AP selection solutions in other networks with different numbers of APs or UEs. Even under the same number of APs or UEs, a trained agent for a given urban network is highly unlikely to perform well in other networks and requires re-training, since the propagation in different urban areas is likely to be highly dissimilar. To illustrate this, Fig. \ref{DRLcompare} shows the throughput performance of the \textit{UnifSrv-DRL} agent trained for Seoul in the Frankfurt network with $M=324$, which is uniformly subsampled to match the number of APs in Seoul, and \mbox{vice-versa}. For both Seoul and Frankfurt, the agent trained for one environment performs poorly in another, due to the significant difference in the underlying propagation environment (\textit{cf. }\mbox{Fig. \ref{snr}}). Fig. \ref{DRLcompare} thus illustrates the limitation in the generalization ability of \mbox{\textit{UnifSrv-DRL}}, incurring high re-training complexity when applying to different network deployments and propagation. We note that this behavior is not specific to our algorithm, but rather reflects a general limitation of \mbox{DRL-based} algorithms \cite{DRLmobi,drlJP,DRLmaxRate, DRLmaxMinRate}, including the \textit{PF-DRL} benchmark, which can be improved via techniques such as transfer-learning but at additional cost. In contrast to the DRL-based algorithm \textit{UnifSrv-DRL}, our heuristic \textit{UnifSrv-heu} algorithm requires no training nor tuning of parameters, since the threshold $\alpha$ is fully dynamic. \textit{UnifSrv-heu} can thus be directly applied to any urban topology and propagation environment, like the other heuristic benchmarks (\textit{PUC}, \mbox{\textit{PUC-const}}, and \textit{CUC}). Our \textit{UnifSrv-heu} algorithm thus has significantly better generalization ability than \mbox{\textit{UnifSrv-DRL}}, and is thus more suitable for practical deployments.\looseness=-1

\subsubsection{Design of Threshold $\alpha$ for \textit{UnifSrv-heu}}

\begin{figure}[!tb]
	\centering
	\subfigure[Stablized $\alpha$ vs. channel realization]{
	     \label{HEUcompare.alpha}
	     \includegraphics[width=0.5\columnwidth]{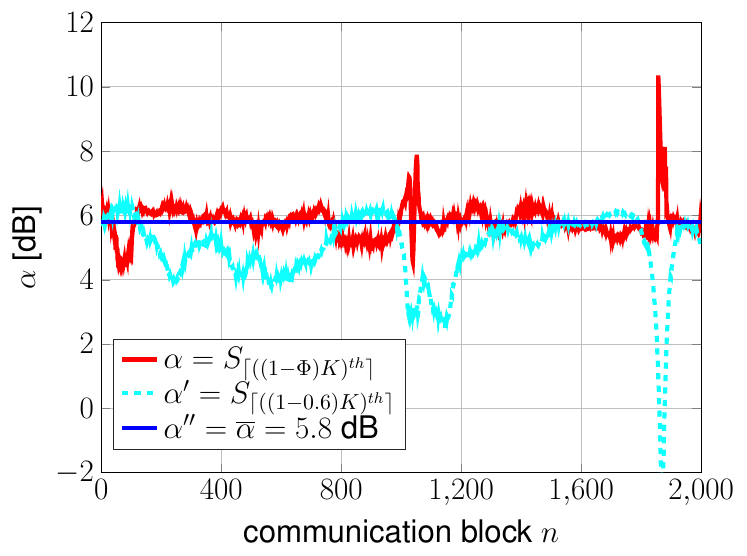}}
    \subfigure[spectral efficiency]{
	     \label{HEUcompare.SE}
	     \includegraphics[width=0.45\columnwidth]{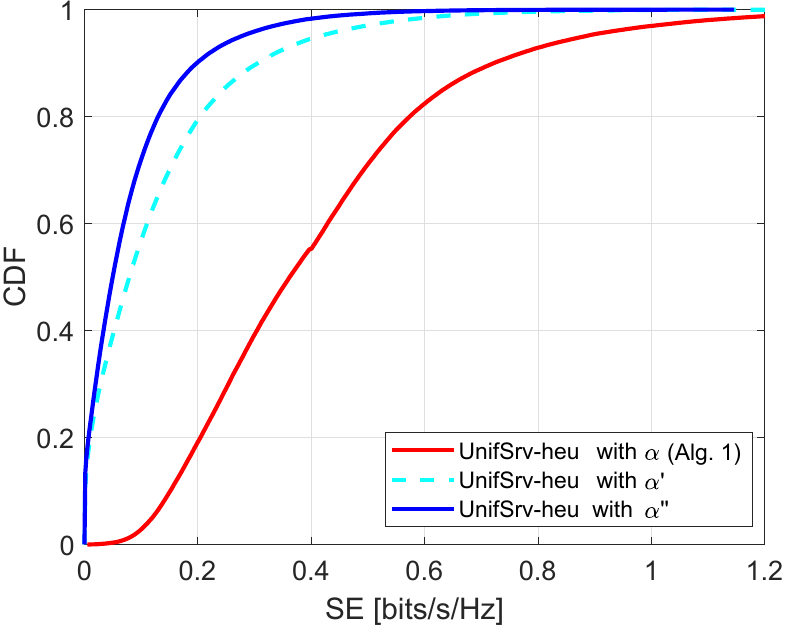}}
	\caption{The value of threshold $\alpha$, fairness index $\Phi$, and spectral efficiency of \textit{UnifSrv-heu} with different threshold designs in Seoul, with $M=324$ APs ($\lambda = 576 \text{ AP/km}^2$).}
	\label{HEUcompare}
\end{figure}

In Alg. \ref{partialFair}, we obtain $\alpha$ via Jain's fairness index $\Phi$ to maximize the throughput fairness of all UEs, in line with objective (\ref{objFair}) of our optimization problem. To demonstrate the validity of our design of $\alpha$ as a dynamic threshold obtained via $\Phi$, we now consider two variant designs of $\alpha$ in Fig. \ref{HEUcompare} that avoid the calculation of Jain's fairness index $\Phi$ and compare their performance against our design in Seoul (Frankfurt exhibits a similar trend). Firstly, we fix the fairness index used in Step 10 of Alg. \ref{partialFair} to be a constant value of $\Phi=0.6$, which is its time-averaged mean over the entire mobility period obtained via simulation. In this case, we set the threshold to always be the transformed SINR of the $60^\text{th}$ percentile UE, i.e., $\alpha' = S_{\left\lceil ((1-0.6)K)^{th} \right \rceil}$. Secondly, we fix the threshold in Step 10 of Alg. \ref{partialFair} to be $\alpha''=5.8$ dB, as the average of $\alpha$ over the entire mobility period obtained via simulation. Fig. \ref{HEUcompare.SE} shows that the two variant threshold designs obtain consistently and significantly worse throughput performance than our design in \textit{UnifSrv-heu}. Firstly, as Fig. \ref{HEUcompare.alpha} shows, the threshold $\alpha'$ that is variable but obtained via the fixed fairness index in fact exhibits an opposite trend compared with our threshold $\alpha$ based on the instantaneous fairness index. This is because the change of SINR of a fixed-percentile UE follows the same trend as the change in channel conditions. Namely, the SINR of all UEs is low under poor channel conditions, resulting in a low $\alpha'$, preventing more UEs to take more APs into their serving sets, and vice-versa under good channel conditions. The required threshold design should do exactly the opposite: the AP selection algorithm should set a high threshold $\alpha$ to allow UEs to expand their serving set to achieve high serving set quality under bad channel conditions and vice-versa. Furthermore, Fig. \ref{HEUcompare.alpha} shows that the value of $\alpha$ changes significantly, over a range of 6 dB over all communication blocks, demonstrating that the fixed $\alpha''$ cannot reflect the significant difference of channel conditions and accurately identify the UEs that require more serving APs. Only our design of a dynamic threshold $\alpha$ accurately distinguishes the UEs in need of serving APs via the dynamic fairness index, thereby selecting appropriate serving AP sets for all UEs to achieve uniformly high throughput over the network. \looseness=-1

\subsubsection{System Cost}
\label{complex}

\begin{figure}[!tb]
	\centering
	\includegraphics[width=1\columnwidth]{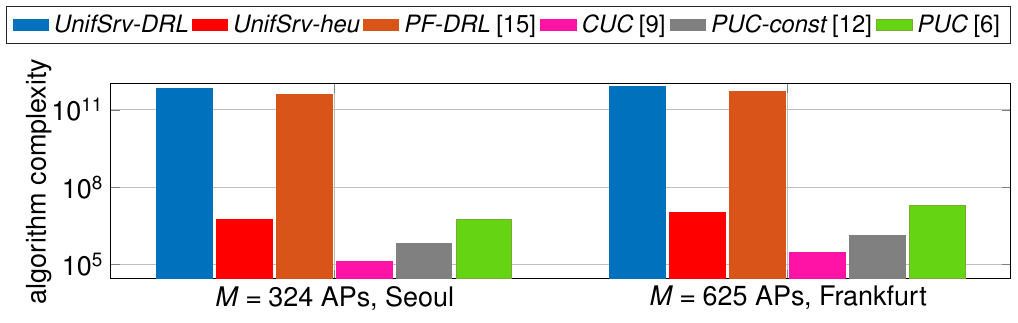}
	\caption{The algorithm complexity of all algorithms in Seoul and Frankfurt.}
	\label{complexUnif}
\end{figure}

Finally, we discuss the system cost of AP selection, in terms of signaling overhead and computational complexity. We firstly note that our \textit{UnifSrv} algorithms do not incur any extra signaling overhead compared to the benchmarks, which comes from transmitting SNR for all UEs as the input of AP selection algorithms and is the same for all considered algorithms. We furthermore compare the computational complexity of all considered AP selection algorithms. For the DQN algorithm underlying \textit{PF-DRL} and \mbox{\textit{UnifSrv-DRL}}, its training complexity is dominated by the number of total training steps and the dimension of hidden layers, given by $\mathcal{O}\left(N_{con} K U_m L n^2\right)$, where $N_{con}$ is the number of episodes to converge, $U_m$ is the number of steps in each round for a UE (i.e., the total number of steps in an episode is $K U_m$), $L$ is the number of hidden layers, and $n$ is the dimension of the hidden layer. We consider training complexity of the DRL-based algorithms in this section, unlike prior works that only consider inference complexity \cite{DRLmobi,drlJP}. This is because, as we showed in Sec. \ref{general}, a trained agent for a given urban \mbox{CF-mMIMO} network is highly unlikely to perform well in other networks and would be required to be re-trained. The training complexity of DRL-based algorithms thus does not lead to a one-time expense but a recurring overhead, and excluding it would understate the true computational burden of deployment in practical \mbox{CF-mMIMO} networks. The complexity of the other algorithms depends on the number of for-loops and the sorting of SNR, which depends on the network scale $M$ and $K$. We give the computational complexity of all considered algorithms in Table \ref{bigO} and present the complexity values in the Seoul and Frankfurt networks in Fig. \ref{complexUnif}. The number of episodes to converge $N_{con}$ is recorded as: $N_{con}=250$ for \textit{UnifSrv-DRL} in Seoul; $N_{con}=300$ for \textit{UnifSrv-DRL} in Frankfurt; $N_{con}=150$ for \textit{PF-DRL} in Seoul; $N_{con}=200$ for \textit{PF-DRL} in Frankfurt. For both cities, \mbox{Fig. \ref{complexUnif}} shows that similarly to \textit{PF-DRL}, \mbox{\textit{UnifSrv-DRL}} entails the highest complexity among all considered algorithms, due to the high dimension of neural network layers and the large number of steps to converge. By contrast, our \textit{UnifSrv-heu} algorithm entails comparable complexity to the heuristic benchmarks, i.e., \textit{CUC}, \mbox{\textit{PUC-const}}, and \textit{PUC}, and is thus recommended for practical urban \mbox{CF-mMIMO} network deployments.\looseness=-1

\begin{table}[!tb]
\caption{Algorithm Complexity.}
\centering
\begin{tabular}{|c|c|}
\hline
\textbf{Algorithm}   & \textbf{Complexity}                          \\ \hline
\textit{UnifSrv-DRL} \& \textit{PF-DRL} \cite{DRLmobi} & $\mathcal{O}\left(N_{con} K U_m L n^2\right)$ \\ \hline
\textit{UnifSrv-heu} (Alg. \ref{partialFair})           & $\mathcal{O}\left( K(MlogM+MlogK+M) \right)$                                    \\ \hline
\textit{PUC} \cite{EE}            & $\mathcal{O}\left( K(MlogM+M^2) \right)$                                    \\ \hline
\textit{PUC-const} \cite{MAcf}           &   $\mathcal{O}\left( K(MlogM+M \tau_p log \tau_p) \right)$                                  \\ \hline
\textit{CUC} \cite{cfsim}           & $\mathcal{O}\left( K(MlogM+E) \right)$                                    \\ \hline
\end{tabular}
\label{bigO}
\end{table}

\section{Conclusions}
\label{conclude}
We conducted a comprehensive performance analysis of CF-mMIMO under realistic urban propagation and proposed the AP selection algorithms \textit{UnifSrv} to achieve uniformly good performance for \mbox{CF-mMIMO} under the highly \mbox{non-uniform} realistic urban environment. We quantified the uniformity of throughput performance via Jain's fairness index and formulated a multi-objective optimization problem to jointly maximize both the sum data rate and the fairness index of the data rate per UE, while minimizing the number of \mbox{AP - UE} connections and thus minimizing the system cost. To solve the optimization problem, we proposed both a DRL and a heuristic AP selection algorithm. We presented extensive simulation results to show that our \textit{UnifSrv} algorithms significantly outperform all considered benchmarks, and thus for the first time achieve both high and uniform throughput of CF-mMIMO in realistic urban environments. Importantly, our heuristic algorithm \mbox{\textit{UnifSrv-heu}} achieves almost equivalent throughput performance to the DRL solution \mbox{\textit{UnifSrv-DRL}}, but with orders of magnitude lower complexity and system cost, making it an attractive solution for practical \mbox{CF-mMIMO}. Our ongoing work is on jointly optimizing the throughput and mobility management cost, i.e., handover overheads, of mobile \mbox{CF-mMIMO} in practical deployments. \looseness=-1

\bibliographystyle{IEEEtran}
\bibliography{IEEEabrv, IEEEexampleMaster.bib}

@INPROCEEDINGS{poster,
  author={Yunlu Xiao and Petri M\"{a}h\"{o}nen and Ljiljana Simi\'{c}},
  booktitle={Proc. IEEE INFOCOM}, 
  title="{Poster abstract: Performance of scalable cell-free
massive MIMO in practical network topologies}", 
  year={2023},
  address = {Hoboken, NJ, USA},
  volume={},
  number={}}

@ARTICLE{EE,
  author={Ngo, Hien Quoc and others},
  journal={IEEE Trans. Green Commun. Netw.}, 
  title="{On the total energy efficiency of cell-free massive MIMO}", 
  year={2018},
  volume={2},
  number={1},
  pages={25-39}
  }

@ARTICLE{green,
  author={Mowla, Md Munjure and Ahmad, Iftekhar and Habibi, Daryoush and Phung, Quoc Viet},
  journal={IEEE Trans. Green Commun. Netw.}, 
  title="{A green communication model for 5G systems}", 
  year={2017},
  volume={1},
  number={3},
  pages={264-280},
  doi={10.1109/TGCN.2017.2700855}}

@article{MIMObook,
year = {2017},
volume = {11},
journal = {Foundations and Trends in Signal Processing},
title="{Massive MIMO networks: spectral, energy, and hardware efficiency}",
number = {3-4},
pages = {154-655},
author = {Emil Bj\"{o}rnson and others}
}

@ARTICLE{SoftHOcf,
  author={Zaher, Mahmoud and Björnson, Emil and Petrova, Marina},
  journal={IEEE Trans. Wireless Commun.}, 
  title = "{Soft handover procedures in mmWave cell-free massive MIMO networks}",
  year={2024},
  volume={23},
  number={6},
  pages={6124-6138}
  }

@ARTICLE{rwp,  author={Lin, Xingqin and others},  journal={IEEE Trans. Wireless Commun.},   title="{Towards understanding the fundamentals of mobility in cellular networks}",   year={2013},  volume={12},  number={4},  pages={1686-1698},  doi={10.1109/TWC.2013.022113.120506}}

@ARTICLE{aging,  author={Zheng, Jiakang and others},  journal={IEEE Trans. Wireless Commun.},   title="{Impact of channel aging on cell-free massive MIMO over spatially correlated channels}",   year={2021},  volume={20},  number={10},  pages={6451-6466},  doi={10.1109/TWC.2021.3074421}}

@book{scboss,
  title="{Small cell networks: deployment, management, and optimization}",
  author={Claussen, Holger and others},
  year={2017},
  publisher={John Wiley \& Sons}
}

@article{cfbook,
    author={Demir, Tugfe and others},
    title = "{Foundations of user-centric cell-free massive MIMO}",
    journal = {Foundations and Trends in Signal Processing},
    volume = {14},
    number = {3-4},
    pages = {162-472},
    year = {2021}}

@INPROCEEDINGS{cfsim,
  author={Interdonato, Giovanni and Frenger, Pal and Larsson, Erik G.},
  booktitle={Proc. IEEE ICC}, 
  title="{Scalability aspects of cell-free massive MIMO}", 
  address = {Shanghai},
  year={2019},
  volume={},
  number={},
  doi={10.1109/ICC.2019.8761828}}

@ARTICLE{cfvs,
  author={Ngo, Hien Quoc and others},
  journal={IEEE Trans. Wireless Commun.}, 
  title="{Cell-free massive MIMO versus small cells}", 
  year={2017},
  volume={16},
  number={3},
  pages={1834-1850},
  doi={10.1109/TWC.2017.2655515}}

@ARTICLE{scalableCF,
  author={Björnson, Emil and Sanguinetti, Luca},
  journal={IEEE Trans. Commun}, 
  title="{Scalable cell-free massive MIMO systems}", 
  year={2020},
  volume={68},
  number={7},
  pages={4247-4261}}

@ARTICLE{MAcf,
  author={Chen, Shuaifei and others},
  journal={China Commun.}, 
  title="{Wireless powered IoE for 6G: Massive access meets scalable cell-free massive MIMO}", 
  year={2020},
  volume={17},
  number={12},
  pages={92-109},
  doi={10.23919/JCC.2020.12.007}}

@ARTICLE{beerten2025,
  author={Beerten, Robbert and others},
  journal={IEEE Open J. Commun. Soc.}, 
  title="{Mobile cell-free massive MIMO: a practical O-RAN-based approach}", 
  year={2025},
  volume={6},
  number={},
  pages={593-610}
  }

@INPROCEEDINGS{shortFI,
  author={Deng, Jing and Han, Yunghsiang S. and Liang, Ben},
  booktitle={Proc. IEEE GLOBECOM}, 
  title="{Fairness index based on variational distance}", 
  year={2009},
  address={Honolulu}}

@ARTICLE{optimalFI,
  author={Sediq, Akram Bin and others},
  journal="{IEEE Trans. Wireless Commun.}", 
  title="{Optimal tradeoff between sum-rate efficiency and Jain's fairness index in resource allocation}", 
  year={2013},
  volume={12},
  number={7},
  pages={3496-3509}
}

@ARTICLE{Cfmeasure,
  author={Löschenbrand, David and others},
  journal={IEEE Access}, 
  title="{Towards cell-free massive MIMO: a measurement-based analysis}", 
  year={2022},
  volume={10},
  number={},
  pages={89232-89247}
  }

@article{2022Evaluation,
  title="{Evaluation of cell-free millimeter-wave massive MIMO systems based on site-specific ray tracing simulations}",
  author={ Silva, P. Higo Thaian Da  and  others},
  journal={IEEE Access},
  volume={10},
  pages={82092-82105},
  year={2022}
}

@INPROCEEDINGS{insite,
  author={Mededovic, Petar and Veletic, Mladen and Blagojevic, Zeljko},
  booktitle={Proc. MIPRO}, 
  title="{Wireless insite software verification via analysis and comparison of simulation and measurement results}", 
  year={2012},
  volume={},
  number={},
  address={Opatija}
  }

@article{viswalk,
title = "{Modeling pedestrian queuing using micro-simulation}",
journal = {Transportation Research Part A: Policy and Practice},
volume = {49},
pages = {232-240},
year = {2013},
author = {Inhi Kim and Ronald Galiza and Luis Ferreira},
}

@INPROCEEDINGS{mywcnc25,
  author={Xiao, Yunlu and Simi\'{c}, Ljiljana},
  booktitle={Proc. IEEE WCNC}, 
  title="{Performance of cell-free massive MIMO in realistic urban propagation environments}", 
  year={2025},
  address = {Milan}
  }

@electronic{shp,
  title         = "{OpenStreetMap data extracts}",
  url           = {https://download.geofabrik.de/},
  year          = {2024}
}

@ARTICLE{scalableFirst,
  author={Buzzi, Stefano and D'Andrea, Carmen},
  journal={IEEE Wireless Commun. Let.}, 
  title="{Cell-free massive MIMO: user-centric approach}", 
  year={2017},
  volume={6},
  number={6},
  pages={706-709}
  }

@ARTICLE{drlJP,
  author={Tsukamoto, Yu and others},
  journal={IEEE Open J. Commun. Soc.}, 
  title="{Scalable AP clustering with deep reinforcement learning for cell-free massive MIMO}", 
  year={2025},
  volume={6},
  number={},
  pages={1552-1567}
}

@ARTICLE{DRLmaxMinRate,
  author={Banerjee, Bitan and others},
  journal={IEEE Trans. Mach. Learn. Commun. Netw.}, 
  title="{Access point clustering in cell-free massive MIMO using conventional and federated multi-agent reinforcement learning}", 
  year={2023},
  volume={1},
  number={},
  pages={107-123}
  }

@ARTICLE{DRLmaxRate,
  author={Gao, Zhichao and others},
  journal={IEEE Commun. Lett.}, 
  title="{DRL-based AP selection in downlink cell-free massive MIMO network with pilot contamination}", 
  year={2024},
  volume={28},
  number={6},
  pages={1432-1436}
}

@ARTICLE{DRLmobi,
  author={Prado, Anna and others},
  journal={IEEE J. Sel. Areas Commun.}, 
  title="{Enabling proportionally-fair mobility management with reinforcement learning in 5G networks}", 
  year={2023},
  volume={41},
  number={6},
  pages={1845-1858}
}

@INPROCEEDINGS{DRLreduceConnect,
  author={Mendoza, Charmae Franchesca and Schwarz, Stefan and Rupp, Markus},
  booktitle={Proc. IEEE ICC}, 
  title="{User-centric clustering in cell-free MIMO networks using deep reinforcement learning}", 
  year={2023},
  address={Rome}
}

@ARTICLE{GMM,
  author={Biswas, Pialy and Mallik, Ranjan K. and Letaief, Khaled B.},
  journal={IEEE Trans. Mach. Learn. Commun. Netw.}, 
  title="{Optimal access point centric clustering for cell-free massive MIMO using Gaussian mixture model clustering}", 
  year={2024},
  volume={2},
  number={},
  pages={675-687}
  }

@ARTICLE{ORAN2025,
  author={Cao, Yang and others},
  journal={IEEE J. Sel. Areas Commun.}, 
  title="{Implementation of a cell-free RAN system with distributed cooperative transceivers under ORAN architecture}", 
  year={2025},
  volume={43},
  number={3},
  pages={765-779}
  }

@ARTICLE{CFmag2026,
  author={Buzzi, Stefano and others},
  journal={IEEE Commun. Mag.}, 
  title="{Why user-centric cell-free distributed MIMO systems will be the disruptive 6G technology}", 
  year={2026},
  volume={},
  number={},
  pages={1-7},
  }

@INPROCEEDINGS{CFfirst,
  author={Ngo, Hien Quoc and others},
  booktitle={Proc. IEEE SPAWC}, 
  title="{Cell-free massive MIMO: uniformly great service for everyone}", 
  year={2015},
  address={Stockholm},
  volume={},
  number={}
  }

@INPROCEEDINGS{Rpilot,
  author={Parida, Priyabrata and Dhillon, Harpreet S.},
  booktitle={IEEE GLOBECOM}, 
  title="{Random sequential adsorption-based pilot assignment for distributed massive MIMO systems}", 
  year={2019},
  address={Waikoloa},
  volume={},
  number={}
  }

@ARTICLE{PUCconstDRL,
  author={Tizikara, Dativa K. and So, Daniel K. C. and Ding, Zhiguo},
  journal={IEEE Trans. Wireless Commun.}, 
  title="{Deep reinforcement learning-based access point selection in cell-free massive MIMO with behavior cloning}", 
  year={2026},
  volume={25},
  number={},
  pages={16229-16244},
  }

\end{document}